\documentclass[aps,prb,twocolumn,superscriptaddress,notitlepage,nofootinbib,longbibliography]{revtex4-1}
\usepackage{mathrsfs}
\usepackage{epsfig}
\usepackage{graphicx}
\usepackage{amsfonts}
\usepackage[figuresright]{rotating}
\usepackage{amssymb}
\usepackage{amsmath}
\usepackage{dcolumn}
\usepackage{bm}
\usepackage{color}
\usepackage{braket}
\usepackage{float}
\usepackage{units}
\usepackage{xspace}

\makeatother
\begin{document}	
%%%%%%%%% title %%%%%%%%%%%
\title{Breakdown of the correspondence between the real-complex and delocalization-localization
transitions in non-Hermitian quasicrystals}
\author{Wen Chen}
\author{Shujie Cheng}
\email{chengsj@zjnu.edu.cn}
\affiliation{Department of Physics, Zhejiang Normal University, Jinhua 321004, China}
\author{Ji Lin}
\affiliation{Department of Physics, Zhejiang Normal University, Jinhua 321004, China}
\author{Reza Asgari}
\affiliation{School of Physics, Institute for Research in Fundamental Sciences (IPM), Tehran 19395-5531, Iran}
\affiliation{Department of Physics, Zhejiang Normal University, Jinhua 321004, China}
\author{Gao Xianlong}
\email{gaoxl@zjnu.edu.cn}
\affiliation{Department of Physics, Zhejiang Normal University, Jinhua 321004, China}

%%%%%%%%% abstract %%%%%%%%

\begin{abstract}
%In a recent research [Phys. Rev. B \textbf{100}, 125157 (2019)], 
The correspondence
between the real-complex transition in energy and delocalization-localization transition is well-established in 
a class of Aubry-Andr\'e-Harper model with exponential non-Hermitian on-site potentials. 
In this paper, we study a generalized Aubry-Andr\'e model with off-diagonal modulation and non-Hermitian 
on-site potential. We find that, when there exists an incommensurate off-diagonal modulation, the correspondence 
breaks down, although the extended phase is maintained in a wide parameter range of the strengths of the 
on-site potential and the off-diagonal hoppings. An additional intermediate phase with a non-Hermitian 
mobility edge emerges when the off-diagonal hoppings become commensurate. This phase is characterized 
by the real and complex sections of the energy spectrum corresponding to the extended and localized states. 
In this case, the aforementioned correspondence reappears due to the recovery of the $\mathcal{PT}$-symmetry.

\end{abstract}
\maketitle

%%%%%%%%% main %%%%%%%%
\section{Introduction}
\label{Introduction}
Anderson localization, an active topic in condensed matter field, tells us that the random disorder fails 
the diffusion of the wave packets, and leads to the localization of the particle \cite{Anderson1958,Abrahams1979,guhr1998random,kramer1993localization}. 
Besides, the quasi-disorder systems, exemplified by Aubry-Andr\'e-Harper (AAH) model~\cite{Aubry1980,Harper1955}, 
are of the similar localized phenomenon but with a transition between extended and localized states with 
the increase of the disordered external potential in a one-dimensional system, which appears only in a three-dimensional 
system of random disorder~\cite{Abrahams1979}. Owing to the rich phase transitions in the quasi-disordered systems, 
there is growing interest both in the theoretical study of the AAH system and its extensions~\cite{Brouwer2013,Chen2012,Giamarchi1999,Zilbergerg2012,Ganeshan2013,Liu2015,zeng2020-1} as well as 
their experimental realizations in the photonic crystals~\cite{Lahini2009,Silberberg2013,Kraus2012,Dal2003} and in 
the ultracold atomic systems~\cite{xiao2021,Roati2008,Modugno2010}.

The AAH model has been extended in numerous ways, such as by include long-range hopping, $p$-wave pairs, and 
off-diagonal modulations, which result in a variety of exotic phenomena. The long-rang hopping term~\cite{Biddle2010} 
or some specific form of the on-site potentials~\cite{Longhi2020} results in the single-particle mobility edges. 
The off-diagonal terms lead to the presence of the critical states in a large parameter space, and brings up phase-transition 
structure of the extended-localized-extended states~\cite{Cestari2016,Liutong2016}. Additionally, the phase diagram 
of the system will change depending on whether the non-diagonal modulation is commensurate or incommensurate.

Further extensions to the AAH model apply to the non-Hermitian systems~\cite{Hatano1996,Longhi2020,Longhi2021,Longhi2021-1,Longhi2021-2,Longhi2019,Longhi2019-1,Bender1998,Chen2017,Chen2017-2,Chen2020,Chen2021,Chen2021-1,Chen2021-2,Chen2021-3,zeng2020-2,PhysRevA.105.032207}, where the corresponding localization, 
mobility edge{\color{red},} and topological properties usually discussed in the Hermitian system are given new properties. The inclusion of nonreciprocal hoppings or complex on-site potentials are theoretically 
the two ways to incorporate non-Hermiticity into disordered systems~\cite{Ueda2019}. Theoretical investigation 
shows that there are synchronous real-complex transition and delocalization-localization transitions in the presence 
of the on-site complex potential~\cite{Longhi2019,Zhang2021}, long-range hoppings~\cite{Longhi2020,Biddle2010,Biddle2011},
and the anti-symmetric hops~\cite{Chen2021,Chen2021-1,Chen2021-2,Chen2021-3}.
The combination of $p$-wave pairings and non-Hermitian quasi-disorder is another important paradigm
to understand the topological properties of quasi-periodic systems~\cite{Liutong2021},
%,Das2010,Kitaev2001,Menke2017,Potter2010,Gao2016,Gao2021,Liutong2017,zeng2020}.
and an unconventional real-complex transition is found in the non-Hermitian quasiperiodic lattice.

We should point out that Longhi~\cite{Longhi2019} systematically examined the relationship between the 
real-complex transition in energy and the delocalization-localization transition in a class of AAH models 
with non-Hermitian exponential potentials and discovered that the extended states correspond to the real 
energies and the localized ones to the complex energies. A key question is whether such a correspondence
is robust against the off-diagonal incommensurate and commensurate hoppings. In this paper, we study a 
generalized AAH model with off-diagonal modulation and non-Hermitian on-site potential and find that, 
for the AAH model with exponential non-Hermitian potential, such a well-established correspondence 
depends on the commensurate off-diagonal hoppings and breaks down when there exists an incommensurate
off-diagonal modulation. 
%REZA: I reckon you should discuss your finding here more.

The rest of this paper is organized as follows. In Sec.~\ref{S2}, we propose the
generalized non-Hermitian AA model with two different types of off-diagonal hoppings.
In Sec.~\ref{S3}, we are devoted to investigating the localization transition
and the properties of the energy spectra under the interplay of the incommensurate modulation and
non-Hermitian on-site potential. In Sec.~\ref{S4}, we study the delocalization-localization 
transition and analyze the properties of the energy spectra under commensurate off-diagonal hoppings.
We make a summary in Sec.~\ref{S5}.

\section{Model and Hamiltonian}
\label{S2}
The interplay of non-Hermiticity and quasidisorder gives rise to new perspectives for the Anderson localization transitions. In this paper, we consider a generalized AAH model  with off-diagonal modulation and non-Hermitian on-site potential. The Hamiltonian is described as
\begin{equation}
	\hat{H}=\sum_{n}^{L-1}\left[t_n\left(\hat{c}_{n+1}^{\dagger}\hat{c}_n+\text{H.C.}\right)\right]+\sum_{n}V_n \hat{c}_n^{\dagger}\hat{c}_n,
	\label{eq1}
\end{equation}
where $L$ is the length of the lattice, $c_n\left(c_n^{\dagger}\right)$ denotes the fermion annihilation (creation) operator at site $n$. The nearest-neighbor hopping amplitude $t_n$ and on-site potential $V_n$ are given by
\begin{equation}
	\begin{split}
		t_n&=t+\lambda\cos(2\pi b_1n+\phi_1),\\
		V_n&=V\exp\left(i(2\pi b_2 n+\phi_2)\right),
	\end{split}
\end{equation}
where $\lambda$ and $V$ denote the modulation amplitudes in the hopping term and on-site complex potential, respectively. An irrational or rational number, $b_1$, is selected, to represent the incommensurate or commensurate potential, respectively.
For the on-site complex incommensurate modulation $b_2$, we discuss in this paper the irrational number $b_2 = (\sqrt{5}-1)/2$.
$\phi_1$ and $\phi_2$ are the extra phases varying from $0$ to $2\pi$. In this work, we set $t$ as the energy unit, and choose $\phi_1=\phi_2\equiv 0$ without loss of generality (Although the phase boundary changes for different values of phases, the main physics keeps the same). 

In the limit of $\lambda = 0$, our model reduces to a non-Hermitian AAH model~\cite{Longhi2019}, which exists a well-defined correspondence between the real-complex transition in energy and delocalization-localization transition, supported by both an analytical and numerical study. Namely, the real energies correspond to the extended states and the complex ones correspond to
the localized states. In the limit of $V = 0$, our model is the one of the limits of the off-diagonal AAH model including both incommensurate and commensurate modulations~\cite{Liutong2016}, which displays extended-critical transition for $b_1=(\sqrt{5}-1)/2$, And regardless of the strength of the modulation, it maintains the extended phases for $b_1=1/2$. 
For the system of the non-Hermitian on-site potential, we are interested in determining whether the delocalization-localization transition still exists and whether the aforementioned connection is resistant to off-diagonal modulations
(See Appendix \ref{app1} for the systems' symmetries).

\section{Incommensurate modulation case}
\label{S3}
First, we study the phase properties of the model with incommensurate modulation $b_1=(\sqrt{5}-1)/2$. The extended, localized, or
critical feature of a specific wave function can be characterized by the inverse participation
ratio (IPR) \cite{PhysRevLett.71.412,PhysRevLett.84.3690} with 
\begin{equation}
    \text{IPR}^{j}=\frac{\sum_{n=1}^L|\psi^{(j)}(n)|^4}{\sum_{n=1}^L|\psi^{(j)}(n)|^2},
\end{equation}
corresponding to $\text{IPR}\to 1$, $\text{IPR}\to0$, and $0<\text{IPR}<1$, 
in the thermodynamic limit, respectively, where $\psi^{(j)}$ means the wave function of the $j$-th eigenstate. Averaging the IPR over all the wave functions, that is, the mean inverse participation ratio (MIPR) with ${\rm MIPR}=\sum_{j=1}^L {\rm IPR}^{(j)}/L$ will be used to analyze the global characteristics of the system.

\begin{figure}[h]
	\centering
	\includegraphics[scale=0.6]{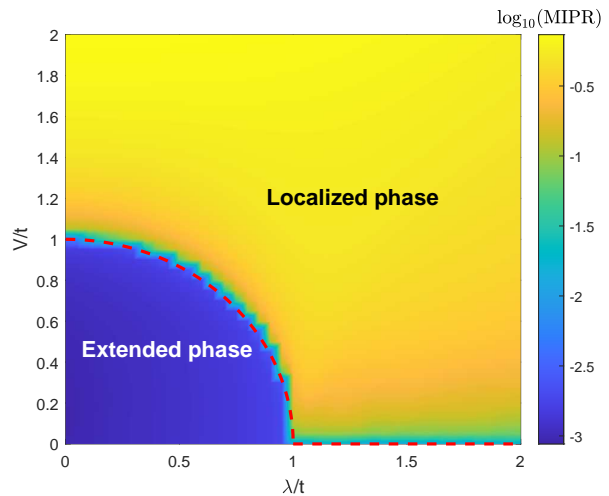}
	\caption{Phase diagram for the incommensurate modulation case presents the $\text{log}_{10}(\rm{MIPR})$ as 
		functions of $V/t$ and $\lambda/t$ with the system size $L=1597$. The red dashed line is the phase boundary based on Eq. (\ref{eq1}),
		separating the extended phase from the localized one. The colorbar is for the value of $\log_{10}$ (MIPR).}
	\label{fig1}
\end{figure}

Having calculated the logarithm of the $\rm{MIPR}$ for a large system size ($L=1597$), we obtain the phase diagram
in the $\lambda$--$V$ parameter space (see Fig.~\ref{fig1}). Intuitively, the phase diagram consists of two different phases separated by the
quarter circle (shown with red dashed line and will be discussed later). In the parameter region surrounded by the inner quarter circle,
the $\text{log}_{10}(\rm MIPR)$ is visibly less than $-1$, meaning that the ${\rm MIPR}$ approaching zero in the large system size and
indicating a delocalized phase. On the contrary, outside the quarter circle, the $\text{log}_{10}(\rm{MIPR})$ tends
to $0$, the $\rm MIPR$ approaching $1$, implying a localized phase. In particular, the $\text{log}_{10}(\rm MIPR)$
is nearly equal to $-1$ on the phase boundary. It means that the wave functions are possible to be critical at these parameter
points. The difference between the phases can be reflected by the spatial distributions of wave functions as well.

\begin{figure}[h]
	\centering
	\includegraphics[scale=0.8]{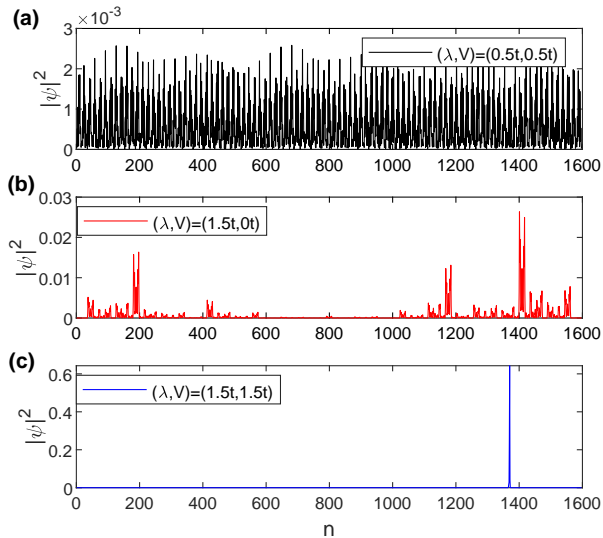}
	\caption{The representative wave functions $\psi_{1000}$. (a)  $(\lambda,V)=(0.5t,0.5t)$ in the extended phase. (b) $(\lambda,V)=(1.5t,0t)$ in the critical phase. (c) $(\lambda,V)=(1.5t,1.5t)$ in the localized phase. The system size is $L=1597$ .}
	\label{fig2}
\end{figure}

In Figs.~\ref{fig2}(a),~\ref{fig2}(b), and~\ref{fig2}(c), we display the spatial distributions of the three representative eigenstates $\psi^{(1000)}$
chosen from the above three phase areas with $(\lambda,V)=(1.5t,1.5t)$, $(1.5t,0t)$, and $(0.5t,0.5t)$, respectively. We see that the wave function is localized
for $(\lambda,V)=(1.5t,1.5t)$, whereas it is extended for $(\lambda,V)=(0.5t,0.5t)$. The wave function is neither
extended nor localized for $(\lambda,V)=(1.5t,0t)$, showing the multi-fractal structure with critical characteristics.
\begin{figure}[h]
	\centering
	\includegraphics[scale=0.8]{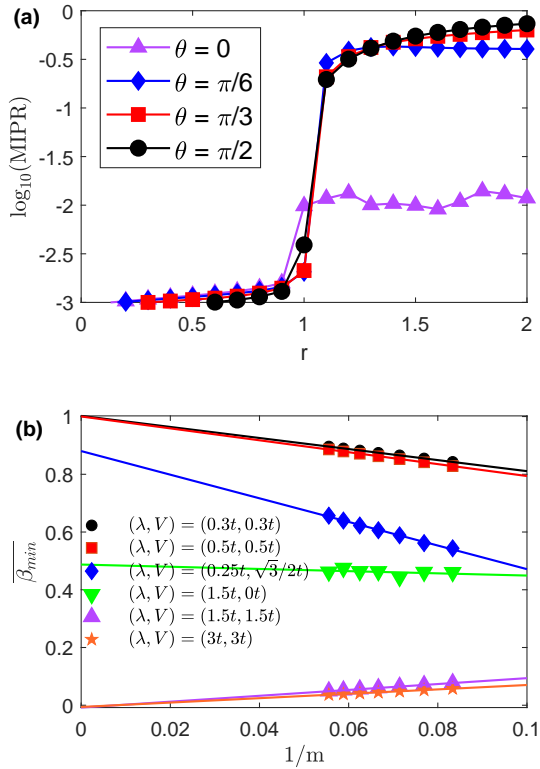}
	\caption{  (a) $\log_{10}$ (MIPR) as a function
		of $r$ with $\theta=0,\pi/6,\pi/3,\pi/2$. (b) $\overline{\beta_{min}}$ as a function of the inverse Fibonacci
		index $1/m$ at different $(\lambda,V)$. Parameter points $(\lambda,V)=(0.3t,0.3t)$ and $(0.5t,0.5t)$ are chosen from
		the extended phase, $(\lambda,V)=(0.5t,\sqrt{3}/2t)$ and $(1.5t,0t)$ are chosen from the critical phase and
		$(\lambda,V)=(1.5t,1.5t), (3t,3t)$ are located in the localized phase.}
	\label{fig3}
\end{figure}

Furthermore, we locate the phase boundary when $\lambda <1$ employing the $\log_{10}$(MIPR). In Fig.~\ref{fig3}(a), we present $\log_{10}(\rm{MIPR})$ as a function
of $r=\sqrt{V^2+\lambda^2}$ for various $\theta={\rm atan}(V/\lambda) \in [0, \pi/2]$ with the system size $L=1597$. The $\log_{10}(\rm{MIPR})$
at various $\theta$ keeps zero in the region $r/t<1$ and are finite numbers when $r/t>1$, showing a phase transition at $r/t=1$,
namely $\sqrt{V^2+\lambda^2}=t$. Particularly, we note that the value of $\log_{10}(\rm{MIPR})$ at $\theta=0$ for $r/t>1$ (corresponding to $V=0$ and $\lambda>1$)
is significantly different from those for other $\theta$s. The slightly smaller finite value of $\log_{10}(\rm{MIPR})$ shows the critical characteristic of the system
at these parameter points. The results are self-consistent with the fractional dimension discussed below. From the above analysis, the complete phase boundary is
described by
\begin{equation}
	\label{eq4}
	\begin{cases}
		V^2+\lambda^2=t^2,\quad &V\neq 0\\
		\lambda>t,&V=0.
	\end{cases}
\end{equation}
However, in the axis of ordinates, for $\lambda = 0t$ and $V>t$, the system is in the localized phase. As a result, the phase boundary $V^2+\lambda^2=t^2$ in Fig.~\ref{fig1} by the red dashed line is due to the numerical simulations. The exact phase boundary in the present Hamiltonian system is still elusive, while it can be analytically determined in some systems of quasidisorder, for example, for the system of mobility edges by the self-duality condition~\cite{Biddle2010} and for the system of the topological phase transitions by the open and close of the energy gap~\cite{Liu2015}. 

We then use the fractional dimension $\beta$ to validate the arguments made previously regarding the various phases.  For a system with $L=F_{m}$ ($F_m$ is the $m$-th Fibonacci number),
the fractional dimension at lattice site $n$, i.e., $\beta_n$ can be extracted from
\begin{equation}\label{beta_n}
p_n=F^{-\beta_{n}}_m,
\end{equation}
where $p_n$ is the probability density. From the above equation, we know that this quantity $\beta_{n}$ plays the role of a scaling index. $\beta_n\sim 1$ for an extended state since $p_n\sim 1/F_m$. For a localized state, $\beta_n\sim 1$ on those localized sites and $\beta_n\rightarrow \infty$ on the other unoccupied sites. For a critical state, the index $\beta_n$ is within a finite interval $[\beta^{n}_{min},\beta^{n}_{max}]$. As a result, the minimal $\beta_n$, i.e., $\beta^{n}_{min}$, is a direct feedback of the characteristic of a designated wave function~\cite{Gao2016}. Specifically,
$\beta^{n}_{min} \rightarrow 0$ signals a localized state, $0< \beta^{n}_{min}< 1$ a critical state, and $\beta^{n}_{min} \rightarrow 1$ an extended state. Without loss of generality, we employ the average of $\beta^{n}_{min}$ over all states, i.e. $\overline{\beta_{min}}=\sum_{i}\beta^{i}_{min}/L$
under the extrapolation limit $1/m \rightarrow 0$ to distinguish different phases. We choose some typical parameter points in
various phases to calculate the $\overline{\beta_{min}}$. As shown in Fig.~\ref{fig3}(b), we find that the corresponding $\overline{\beta_{\text{min}}}$
both tends to $1$ at $(\lambda,V)=(0.3t,0.3t)$ and $(0.5t,0.5t)$, verifying that the system is in the extended phase. As predicted,  the corresponding
$\overline{\beta_{\text{min}}}$ both approach the value within $(0,1)$ in the thermodynamic limit at $(\lambda,V)=(0.25t,\sqrt{3}/2t)$
and $(1.5t,0t,)$, showing the distinctly critical characteristics. At $(\lambda,V)=(1.5t,1.5t)$ and $(3t,3t)$, the corresponding $\overline{\beta_{\text{min}}}$
extrapolates to $0$, identifying that the system is in the localized phase.

\begin{figure}[h]
	\centering
	\includegraphics[scale=0.6]{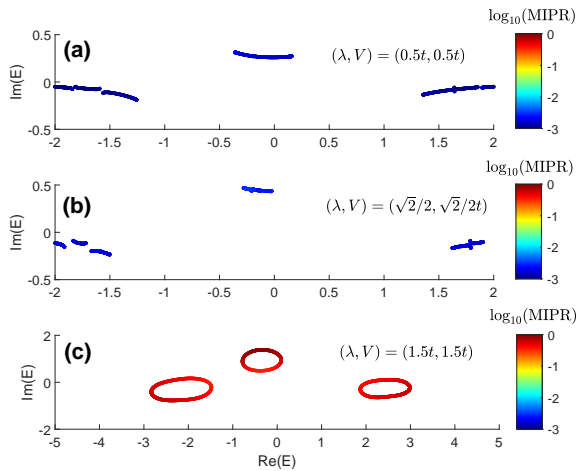}
	\caption{Real energy spectrum of the incommensurate modulation case shown in the complex plane with (a) $(\lambda,V) = (0.5t,0.5t)$; (b) $(\lambda,V) = (\sqrt{2}/2t,\sqrt{2}/2t)$; and (c) $(\lambda,V) = (1.5t,1.5t)$. The colorbar is for the value of $\log_{10}$ (MIPR). The system size is $L=1597$.}
	\label{fig4}
\end{figure}

\begin{figure}[h]
	\centering
	\includegraphics[scale=0.6]{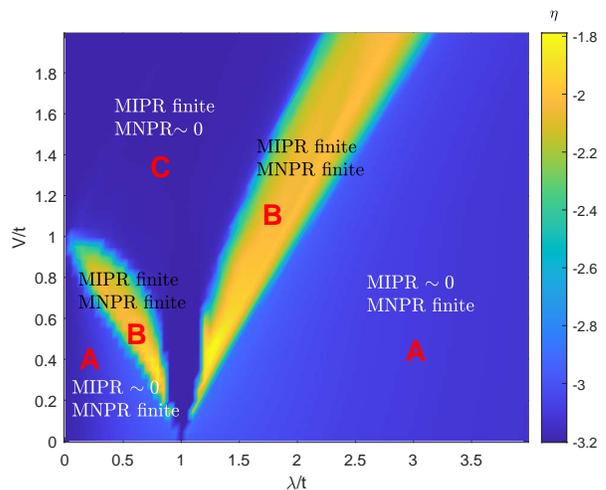}
	\caption{Phase diagram of the commensurate modulation case presents $\eta$ as a function of the parameters $\lambda$ and $V$ with the system size $L=1597$. There are three different phases.
		Region B is the intermediate phase with $-2\leq \eta \leq -1$, where both $\text{MIPR}$ and $\text{MNPR}$ are finite, and Im$(E)\neq 0$. Region A is surrounded
		by the Region B and X-axis and indicates extended phase and pure real energy spectrum with $\eta<-3$, where MIPR is finite and $\text{MNPR}\sim 0$, and Im$(E)\neq 0$.
		The rest of the phases diagram is Reigion C, where energy spectrum is complex and wave function is localized with $\eta<-3$. $\text{MIPR}\sim 0$ and MNPR is finite, and Im$(E)=0$. The colorbar is for the value of $\eta$.}
	\label{fig5}
\end{figure}

 In Ref.~\cite{Longhi2019}, there is a strict correspondence between the delocalization-localization transition and
the real-complex transition, and energy spectra in complex plane is an ellipse symmetric about $\text{Im}(E)=0$, which means that the complex energies come in the form of conjugate pairs. The reason for this phenomenon is that the Hamiltonian is of $\mathcal{PT}$-symmetric and as the strength of the non-Hermitian term changes, it undergoes a transition from a $\mathcal{PT}$-symmetric phase to a $\mathcal{PT}$-broken one. The off-diagonal incommensurate hopping modulation studied in this paper, breaks the $\mathcal{PT}$ symmetry. Hence, such a correspondence may not exist. Taking parameter points $(\lambda,V) = (0.5t,0.5t)$, $(\sqrt{2}/2t,\sqrt{2}/2t)$, and $(1.5t,1.5t)$ corresponding to the extended phase, the critical phase at the boundary, and the localized phase. The corresponding energy spectra are plotted in Figs.~\ref{fig4}(a),~\ref{fig4}(b) and ~\ref{fig4}(c), respectively. Intuitively, all energy spectra are complex. Thus, there is no real-complex transition of the energy spectra accompanying the delocalization-localization phase transition. Moreover, the energy spectrum on the complex plane is not symmetric about $\text{Im}(E)=0$, due to the destruction of the $\mathcal{PT}$-symmetry. As a result, the strict correspondence between the delocalization-localization transition and
the real-complex transition is not robust and broken by the off-diagonal incommensurate hopping modulation. 

Besides the broken correspondence between the delocalization-localization transition and the real-complex transition, we notice that the energy spectra form open arms both in the extended and critical phases, while in the localized phases, the energy spectra remain the closed loops. This situation is quite different from the present results in the literature\cite{Longhi2019,Longhi2019-1}. In these paper, there is a conventional correspondence among arc-loop shape, real-complex spectra, and delocalized-localized transitions. Although the real-complex transition is broken down, the correspondence between arc-loop shape and real-complex spectra is still preserved.

\section{Commensurate modulation case}
\label{S4}
Now we focus on the commensurate modulation case. Without loss of generality, we choose $b_1=1/2$. 
The novel effective quantity $\eta$, suggested by Li et al. ~\cite{Lixiao2020}, is used in the numerical analyses and given by
\begin{equation}
	\label{eq3}
	\eta = \log_{10}[\text{MIPR}\times \text{MNPR}],
\end{equation}
where MNPR is the abbreviation for the mean of the normalized participation ratio which is the normalized participation ratio (NPR) averaged over all eigenstates ${\rm MNPR}=\sum_{j=1}^L {\rm NPR}^{(j)}/L$ with ${\rm NPR}^{(j)}$ defined as,
\begin{equation}
	\text{NPR}^{j}=\left[L\frac{\sum_{n=1}^L|\psi^{(j)}(n)|^4}{\sum_{n=1}^L|\psi^{(j)}(n)|^2}\right]^{-1},
\end{equation}
used for separating the intermediate phase from the extended and localized ones.
In the intermediate phase, both quantities of $\text{MIPR}$ and $\text{MNPR}$ are finite
$[\sim \mathcal{O}(1)]$, leading to $-2\leq \eta \leq -1$. In the extended or localized phase, one of the two quantities
scales as $\sim L^{-1}$, leading to $\eta < -\log_{10}L\sim -3$. By calculating $\eta$, the full phase diagram of the commensurate modulation case with $L=1597$ ($\sim 10^{3}$) is depicted in Fig.~\ref{fig5}. Intuitively, the phase
diagram consists of two main regions. The region highlighted in blue, is marked by $\rm{A}$ or $\rm{C}$, with $\eta$ less than $-3$, which is further distinguished by $\text{MNPR}$ and $\text{MIPR}$.
The other region has the value of $\eta$ slightly larger: $-2 \leq\eta \leq -1$, marked by $\rm{B}$, corresponding to the intermediate phase.

\begin{figure}[h]
	\centering
	\includegraphics[scale=0.7]{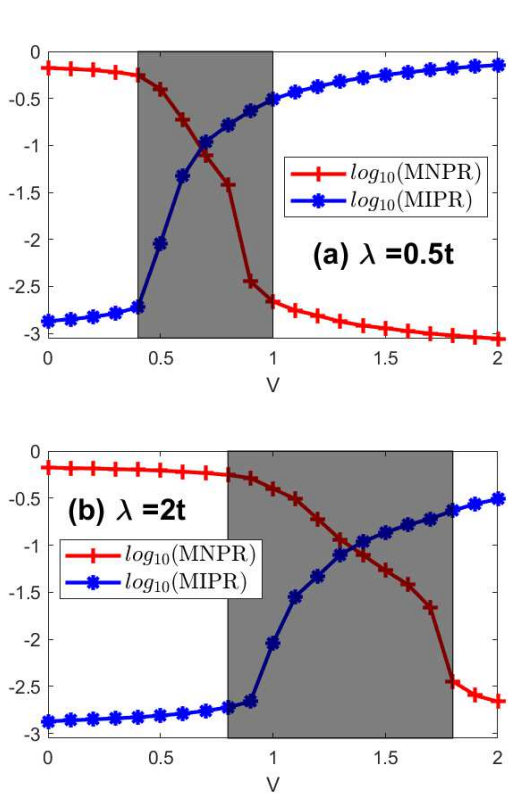}
	\caption{$\log_{10}$($\text{MNPR}$) and $\log_{10}$($\text{MIPR}$) versus $V/t$ in the case of the commensurate modulation $b_1=1/2$ for (a) $\lambda=0.5t$ and (b) $2t$. The shaded area shows the intermediate phases. The system size is $L=1597$. }
	\label{fig6}
\end{figure}

Now we further distinguish the region $\rm{A}$ and $\rm{C}$. Taking $\lambda=0.5$
as an example, $\text{log}_{10}(\text{MNPR})$ and $\text{log}_{10}(\text{MNPR})$ as a function of the potential
strength $V$ are plotted in Fig.~\ref{fig6}. Clearly, as $V$ increases, the system undergoes the extended ($\text{MNPR}\sim \mathcal{O}(1)$
and $\text{NIPR}\sim L^{-1}$), intermediate ($\text{MNPR}\sim \mathcal{O}(1)$
and $\text{MIPR}\sim\mathcal{O}(1)$), and localized ($\text{MNPR}\sim L^{-1}$ \
and $\text{MIPR}\sim\mathcal{O}(1)$) phases. With this distinct definition, we make it clear that the blue region
in Fig.~\ref{fig5} consists of the extended phase (marked by $\rm{A}$) and the localized phase (marked by $\rm{C}$).

\begin{figure}[h]
	\centering
	\includegraphics[scale=0.45]{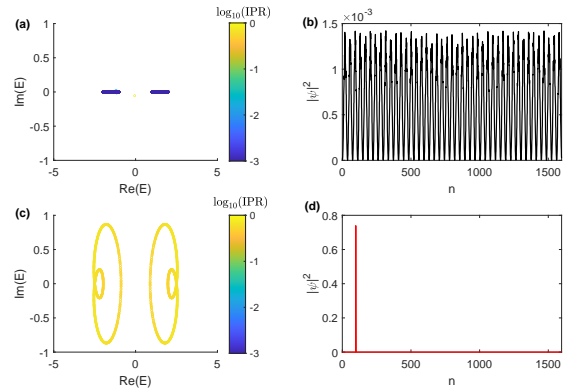}
	\caption{{\bf (a)} Real energy spectrum of the commensurate modulation case shown in the complex plane
		with  $(\lambda,V) = (0.5t,0.1t)$. The colorbar is $\log_{10}$(IPR). {\bf (b)} Spatial distribution of the
		wave function $\psi^{(100)}$ at $(\lambda,V) = (0.5t,0.1t)$. {\bf (c)} Complex energy spectrum shown in the complex plane
		at $(\lambda,V) = (0.5t,1t)$. The colorbar is $\log_{10}$(IPR). {\bf (d)} Spatial distribution of the wave function
		$\psi^{(100)}$ at $(\lambda,V) = (0.5t,1t)$.  The colorbar is for the value of $\log_{10}$ (IPR). Other involved parameter is $L=1597$.}
	\label{fig7}
\end{figure}

In the case of the commensurate modulation, with the recovery of the $\mathcal{PT}$-symmetry, we find that the correspondence between the delocalization-localization phase transition
and the real-complex transition in energy reappears. Taking the parameter point $(\lambda,V)=(0.5t,0.1t)$ in the extended phase
as an example, the energy spectrum is plotted in Fig.~\ref{fig7}(a), from which we see that the energies are all real with
$\text{log}_{10}(\rm{IPR}) \sim -3$. Fig.~\ref{fig7}(b) is the density distribution of the extended wave
function $\psi^{(100)}$. On the contrary, at the parameter point $(\lambda,V)=(1t,0.5t)$ taken from the localized phase,
the corresponding energies are complex with $\text{log}_{10}(\rm{IPR})\sim 0$, shown in Fig.~\ref{fig7}(c). Fig.~\ref{fig7}(d)
presents the localized density distribution of the wave function $\psi^{(100)}$.

\begin{figure}[h]
	\centering
	\includegraphics[scale=0.6]{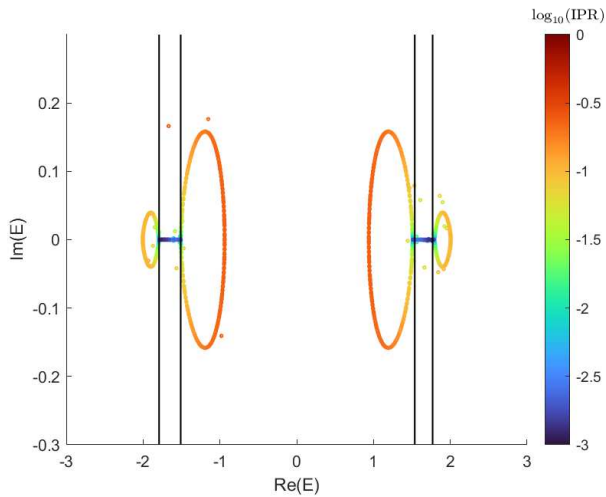}
	\caption{Energy spectrum of the commensurate modulation case plotted in the
		complex plane with $(\lambda,V) = (0.5t,0.7t)$ and $L=1597$. The black dashed lines are mobility edges. The colorbar shows the value of $\log_{10}$(IPR). }
	\label{fig8}
\end{figure}

Meanwhile, in the intermediate phase, the aforementioned correspondence still exists. Fig.~\ref{fig8} presents
the energy spectrum under $(\lambda,V)=(0.5t,0.7t)$ (taken from the intermediate phase) with $L=1597$. Intuitively,
the energies of the extended states with $\text{log}_{10}(\rm IPR) \sim -3$ are all real and the ones of the localized states
with $\text{log}_{10}(\rm IPR)$ larger than $-3$ are fully complex. Thus, in this situation, we obtain the non-Hermitian mobility edges, characterized by the real and complex parts of the energy spectrum corresponding to the extended and localized states, respectively. And because of the $\mathcal{PT}$-symmetry (See Appendix \ref{app1}), complex energies always appear in the form of conjugate pairs (also as in Fig.~\ref{fig7}(c)) and the appearance of the real energy is associated with the mobility edge.

\begin{figure}
    \centering
    \includegraphics[scale=0.6]{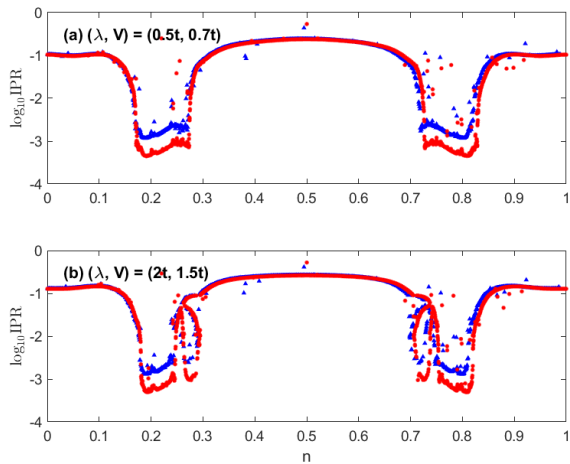}
    \caption{Distributions of the IPRs for the systems with (a) $(\lambda,V) = (0.5t,0.7t)$ and  (b) $(\lambda,V) = (2t,1.5t)$. Red points and blue triangles correspond to the results under different system size, $L = 1597$ and $L = 4181$.}
    \label{fig10}
\end{figure}

\begin{figure}[h]
	\centering
	\includegraphics[scale=0.6]{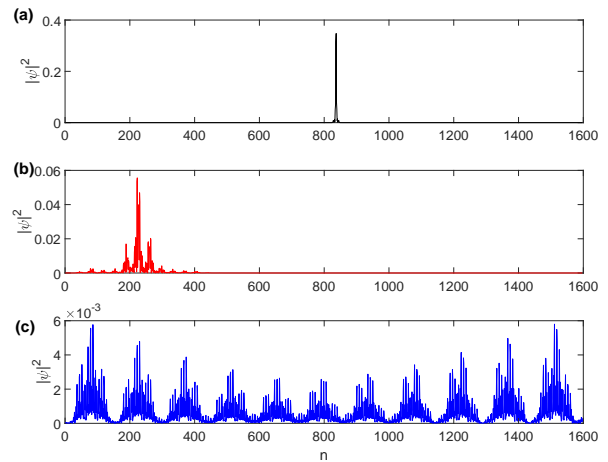}
	\caption{The representative wave functions for the systems with $(\lambda,V) = (0.5t,0.7t)$ and $L=1597$. (a) $\rm{Re(E)}\approx 0.94$; (b) $\rm{Re(E)}\approx 1.51$;
		(c) $\rm{Re(E)}\approx 1.62$, corresponding to the eigenstates in the complex plane in Fig. \ref{fig8}.}
	\label{fig9}
\end{figure}

To further explore the details of the mobility edge, we present the distributions of IPRs for the systems with $(\lambda,V) = (0.5t,0.7t)$ and $(\lambda,V) = (2t,1.5t)$ in Fig.~\ref{fig10}. The states are arranged in the ascending order of the real parts of the energies. In the mixed phase, the IPR varies from a finite value to $1/L$, and the sudden changes in the distributions of the IPRs indicate the presence of a mobility edge. In Fig. \ref{fig9}, we show the representative spatial density distributions for the systems with $(\lambda,V) = (0.5t,0.7t)$ and $L=1597$, and the corresponding energy spectra are plotted in Fig.~\ref{fig8}. Fig. \ref{fig9}(a) shows the localized states with complex energy and its real part ${\rm Re(E)}\sim 0.94t$ (taken from the bigger energy loop in Fig. \ref{fig8}) and \ref{fig9}(b) the localized states with ${\rm Re(E)}\sim 1.51t$
(the smaller energy loop in Fig. \ref{fig8}). Fig.~\ref{fig9}(c) shows one of the extended state with real energy $E\sim 1.62t$ (corresponding to the real energy spectra in Fig. \ref{fig8}).

\section{Summary}
\label{S5}
In this work, we have numerically studied a generalized Aubry-Andr\'e-Harper model with off-diagonal incommensurate or commensurate hoppings and non-Hermitian exponential potential. The correspondence between the real-complex transition and delocalization-localization
transition is missing due to the incommensurate off-diagonal modulations, which breaks the  $\mathcal{PT}$-symmetry. Besides, we find that for the incommensurate hopping, the phase diagram shows that
the extended phase and localized one are separated by the phase boundary $V^2+\lambda^2=t^2$. The extended
phase forms a quarter-circle area in the positive definite parameter region. The extended-localized transition is
self-consistently analyzed by the MIPR and fractal dimension. For the commensurate hopping case, there appears
an extra intermediate phase, which has the non-Hermitian mobility edge, which splits the real and complex parts of the energy spectrum and simultaneously the extended and localized states. Due to the recovery of the $\mathcal{PT}$-symmetry, separated by the mobility edge, we find that the correspondence between the real-complex transition in energy and the delocalization-localization phase transition reappears. 

We notice that a generalized AAH model with both diagonal and off-diagonal quasi-periodic disorders is recently realized by the technique of momentum-lattice engineering in the ultracold atomic system~\cite{xiao2021,Liutong2016}, 
where the topological phase with the critical localization in a quasi-periodic lattice is observed. In the experimental implementation of 
Ref. \cite{xiao2021}, $^{87}{\rm Rb}$ atoms are confined in a one-dimensional momentum lattice. Due to a series of 
frequency-modulated Bragg-laser pairs, the discrete momentum states of the atoms are coupled together. 
Once the Bragg-coupling parameters between nearest-neighbor sites are adjusted, both the diagonal and 
off-diagonal disorders are imposed. In addition, the complex quasi-periodic on-site potential can be realized 
by a low-finesse intracavity etalon, with free spectral range incommensurate with respect to the modulation 
frequency and much smaller than the gain bandwidth \cite{Longhi2019}. Besides, an excellent experimental progress has been made in the observation of the topological triple phase transition in non-Hermitian Floquet quasicrystals via photonic quantum walks\cite{weidemann2022topological}. As a result, we believe that the breakdown of the correspondence between the real-complex and delocalization-localization transitions for the model we studied here 
is of the potential to be realized in the cold atomic experiments or in the photonic quantum walks.
\section*{Acknowledgments}
The authors acknowledge support from NSFC under
Grants No. 11835011 and No. 12174346. R. A. thanks R. R. Tabar for fruitful discussions. 

%%%%%%%%% appendix %%%%%%%%
\appendix
\section{Symmetries of the system}
\label{app1}
The effects of parity operator $\hat{\mathcal{P}}$ and time-reversal operator $\hat{\mathcal{T}}$ on discrete systems are as follows,
\begin{equation}
 \hat{\mathcal{P}}\hat{c}^{\dagger}_n \hat{\mathcal{P}}=\hat{c}^{\dagger}_{L+1-n},\quad  \hat{\mathcal{T}}i\hat{\mathcal{T}}=-i.
\end{equation}
And the Hamiltonian (see Eq.~(\ref{eq4})) can be expressed as follow, 
\begin{equation}
    \hat{H} = \hat{H}_{\rm hopping}+\hat{H}_{\rm{potential}}.
\end{equation}
At first, we consider the non-Hermitian quasiperiodic potential,
\begin{equation}
    \hat{H}_{\rm{potential}} = \sum_n^L \exp(i(2\pi b_2n+\phi_2))\hat{c}_n^{\dagger}\hat{c}_n.
\end{equation}

Under the $\hat{\mathcal{P}}\hat{\mathcal{T}}$ operator,
\begin{equation}
    \begin{split}
        &\hat{\mathcal{P}}\hat{\mathcal{T}}\hat{H}_{\rm{potential}}\hat{\mathcal{T}}\hat{\mathcal{P}}\\
        =& \hat{\mathcal{P}}\hat{\mathcal{T}}\sum_n^L \exp(i(2\pi b_2n+\phi_2))\hat{c}_n^{\dagger}\hat{c}_n \hat{\mathcal{T}}\hat{\mathcal{P}}\\
        =&\sum_n^L[\cos(2\pi b_2 (L-n+1)-2\pi b_2 (L+1)-\phi_2)+\\
        &i\sin(2\pi b_2 (L-n+1)-2\pi b_2(L+1)-\phi_2)]\hat{c}_{L-n+1}^{\dagger}\hat{c}_{L-n+1}.
    \end{split}
\end{equation}

And then we deal with the hopping terms
\begin{equation}
    \hat{H}_{\rm{hopping}} = \sum_n^{L-1}(t+\lambda\cos(2\pi b_1 n +\phi_1))\left(\hat{c}_{n+1}^{\dagger}\hat{c}_n+\text{H.C.}\right).
\end{equation}
Applying $\hat{\mathcal{P}}\hat{\mathcal{T}}$ operator on hopping terms,
\begin{equation}
    \begin{split}
         &\hat{\mathcal{P}}\hat{\mathcal{T}}\hat{H}_{\rm{hopping}}\hat{\mathcal{T}}\hat{\mathcal{P}}\\
        =&\hat{\mathcal{P}}\hat{\mathcal{T}}\sum_n^{L-1} (t+\lambda\cos(2\pi b_1 n +\phi_1))\left(\hat{c}_{n+1}^{\dagger}\hat{c}_n+\text{H.C.}\right)\hat{\mathcal{T}}\hat{\mathcal{P}}\\
        =& \sum_n^{L-1}(t+\lambda\cos(2\pi b_1 n +\phi_1))\left(\hat{c}_{L-n}^{\dagger}\hat{c}_{L-n+1}+\hat{c}_{L-n+1}^{\dagger}\hat{c}_{L-n}\right)\\
        =& \sum_n^{L-1}(t+\lambda\cos(2\pi b_1 (L-n)-2\pi b_1 L-\phi_1))
        \\&\left(\hat{c}_{L-n}^{\dagger}\hat{c}_{L-n+1}+\hat{c}_{L-n+1}^{\dagger}\hat{c}_{L-n}\right).
    \end{split}
\end{equation}
The $\hat{\mathcal{P}}\hat{\mathcal{T}}$ symmetry is observed,
\begin{equation}
    \hat{\mathcal{P}}\hat{\mathcal{T}} \hat{H}(\phi_1,\phi_2) \hat{\mathcal{T}}\hat{\mathcal{P}}=\hat{H}(\phi_1,\phi_2), 
\end{equation}
when $\tilde{\phi}_1 \equiv -2\pi b_1 L-\phi_1$ and $\tilde{\phi}_2\equiv -2\pi b_2 (L+1)-\phi_2$ satisfy
\begin{equation}
    \tilde{\phi}_1 = -2\pi b_1 L-\phi_1=\phi_1+2k_1\pi,
\end{equation}
\begin{equation}
    \tilde{\phi}_2 = -2\pi b_2 (L+1)-\phi_2 = \phi_2 +2k_2\pi,
\end{equation}
namely,
\begin{equation}
    \phi_1 = -\pi b_1 L -k_1\pi, \quad k_1\in Z,
\end{equation}
\begin{equation}
    \phi_2 = -\pi b_2 (L+1)-k_2\pi,\quad k_2\in Z .
\end{equation}
In this case, the system is invariant under the $\hat{\mathcal{P}}\hat{\mathcal{T}}$ operator. 

In order to ensure that the systems satisfy $\mathcal{P}\mathcal{T}$ symmetry at any phase factor, we need to construct a new anti-unitary operator $\hat{\mathcal{R}}\hat{\mathcal{P}}\hat{\mathcal{T}}$ and eigenenergy spectrum of the system is invariant under the $\hat{\mathcal{R}}$ transformations\cite{Harter2016,Schiffer2021}. If $b_1$ is rational, and the strength of the hopping terms between sites are periodic, we can construct a unitary operator rotating the system by $Mk'$ sites in the counter-clockwise direction,
\begin{equation}
    \hat{\mathcal{R}}_{k'}^{\dagger}\hat{c}_n^{\dagger} \hat{\mathcal{R}}_{k'}=\hat{c}_{n+Mk'}^{\dagger},
\end{equation}
where $M$ is the periodic of the hopping term. It gives
\begin{equation}
    \tilde{\phi}'_1 = \tilde{\phi_1} - \frac{2\pi Mk'}{L}, 
\end{equation}
\begin{equation}
    \tilde{\phi}'_2 = \tilde{\phi_2}-\frac{2\pi Mk'}{L},
\end{equation}
and Hamiltonian under the rotation operators $R_k$ 
\begin{equation}
    \hat{\mathcal{R}}_{k'}^{\dagger}\hat{H}(\tilde{\phi}_1,\tilde{\phi}_2)\hat{\mathcal{R}}_{k'}=\hat{H}(\tilde{\phi}'_1,\tilde{\phi}'_2),
\end{equation}
have the same eigenenergy spectra because $\hat{\mathcal{R}}_k$ is unitary. So if systems have $\hat{\mathcal{P}}\hat{\mathcal{T}}$ symmetry, phase factor $\phi_1$ and $\phi_2$ satisfies
\begin{equation}
    \tilde{\phi}'_1= \tilde{\phi_1} - \frac{2\pi Mk'}{L} = \phi_1+2k_1\pi,
\end{equation}
\begin{equation}
    \tilde{\phi}'_2= \tilde{\phi_2} - \frac{2\pi Mk'}{L} = \phi_2+2k_2\pi,
\end{equation}
namely,
\begin{equation}
    \phi_1 = -\pi b_1 L-k_1\pi+\frac{k'\pi M}{L},\quad k_1,k'\in Z,
\end{equation}
\begin{equation}
    \phi_2 = -\pi b_2 (L+1) - k_2\pi +\frac{k'\pi M}{L},\quad k_2,k' \in Z,
\end{equation}
the number of points satisfy condition above becomes infinite for $L\to \infty$. So in the thermodynamic limit, if $b_1$ is rational, systems have $\hat{\mathcal{P}}\hat{\mathcal{T}}$ symmetry.  On the other hand, if $b_1$ is irrational, the systems do not have rotation symmetry so the $\hat{\mathcal{P}}\hat{\mathcal{T}}$ symmetry is broken.
%The non-Hermitian quasiperiodic potential in our model (see Eq.~(\ref{eq1})) is invariant under $\hat{\mathcal{P}}\hat{\mathcal{T}}$ operator because it satisfies 
%\begin{equation}
%    V_n=V^*_{-n},\quad n=1,2,\dots,L
%\end{equation}
%for $\phi_2=0$. 
%%%%%%%%% reference %%%%%%%
\bibliography{ref}

\begin{thebibliography}{51}%
\makeatletter
\providecommand \@ifxundefined [1]{%
 \@ifx{#1\undefined}
}%
\providecommand \@ifnum [1]{%
 \ifnum #1\expandafter \@firstoftwo
 \else \expandafter \@secondoftwo
 \fi
}%
\providecommand \@ifx [1]{%
 \ifx #1\expandafter \@firstoftwo
 \else \expandafter \@secondoftwo
 \fi
}%
\providecommand \natexlab [1]{#1}%
\providecommand \enquote  [1]{``#1''}%
\providecommand \bibnamefont  [1]{#1}%
\providecommand \bibfnamefont [1]{#1}%
\providecommand \citenamefont [1]{#1}%
\providecommand \href@noop [0]{\@secondoftwo}%
\providecommand \href [0]{\begingroup \@sanitize@url \@href}%
\providecommand \@href[1]{\@@startlink{#1}\@@href}%
\providecommand \@@href[1]{\endgroup#1\@@endlink}%
\providecommand \@sanitize@url [0]{\catcode `\\12\catcode `\$12\catcode
  `\&12\catcode `\#12\catcode `\^12\catcode `\_12\catcode `\%12\relax}%
\providecommand \@@startlink[1]{}%
\providecommand \@@endlink[0]{}%
\providecommand \url  [0]{\begingroup\@sanitize@url \@url }%
\providecommand \@url [1]{\endgroup\@href {#1}{\urlprefix }}%
\providecommand \urlprefix  [0]{URL }%
\providecommand \Eprint [0]{\href }%
\providecommand \doibase [0]{http://dx.doi.org/}%
\providecommand \selectlanguage [0]{\@gobble}%
\providecommand \bibinfo  [0]{\@secondoftwo}%
\providecommand \bibfield  [0]{\@secondoftwo}%
\providecommand \translation [1]{[#1]}%
\providecommand \BibitemOpen [0]{}%
\providecommand \bibitemStop [0]{}%
\providecommand \bibitemNoStop [0]{.\EOS\space}%
\providecommand \EOS [0]{\spacefactor3000\relax}%
\providecommand \BibitemShut  [1]{\csname bibitem#1\endcsname}%
\let\auto@bib@innerbib\@empty
%</preamble>
\bibitem [{\citenamefont {Anderson}(1958)}]{Anderson1958}%
  \BibitemOpen
  \bibfield  {author} {\bibinfo {author} {\bibfnamefont {P.~W.}\ \bibnamefont
  {Anderson}},\ }\bibfield  {title} {\enquote {\bibinfo {title} {{Absence of
  diffusion in certain random lattices}},}\ }\href {\doibase
  10.1103/PhysRev.109.1492} {\bibfield  {journal} {\bibinfo  {journal} {Phy.
  Rev.}\ }\textbf {\bibinfo {volume} {109}},\ \bibinfo {pages} {1492} (\bibinfo
  {year} {1958})}\BibitemShut {NoStop}%
\bibitem [{\citenamefont {Abrahams}\ \emph {et~al.}(1979)\citenamefont
  {Abrahams}, \citenamefont {Anderson}, \citenamefont {Licciardello},\ and\
  \citenamefont {Ramakrishnan}}]{Abrahams1979}%
  \BibitemOpen
  \bibfield  {author} {\bibinfo {author} {\bibfnamefont {E.}~\bibnamefont
  {Abrahams}}, \bibinfo {author} {\bibfnamefont {P.~W.}\ \bibnamefont
  {Anderson}}, \bibinfo {author} {\bibfnamefont {D.~C.}\ \bibnamefont
  {Licciardello}}, \ and\ \bibinfo {author} {\bibfnamefont {T.~V.}\
  \bibnamefont {Ramakrishnan}},\ }\bibfield  {title} {\enquote {\bibinfo
  {title} {{Scaling theory of localization: Absence of quantum diffusion in two
  dimensions}},}\ }\href {\doibase 10.1103/PhysRevLett.42.673} {\bibfield
  {journal} {\bibinfo  {journal} {Phy. Rev. Lett.}\ }\textbf {\bibinfo {volume}
  {42}},\ \bibinfo {pages} {673} (\bibinfo {year} {1979})}\BibitemShut
  {NoStop}%
\bibitem [{\citenamefont {Guhr}\ \emph {et~al.}(1998)\citenamefont {Guhr},
  \citenamefont {M{\"u}ller-Groeling},\ and\ \citenamefont
  {Weidenm{\"u}ller}}]{guhr1998random}%
  \BibitemOpen
  \bibfield  {author} {\bibinfo {author} {\bibfnamefont {Thomas}\ \bibnamefont
  {Guhr}}, \bibinfo {author} {\bibfnamefont {Axel}\ \bibnamefont
  {M{\"u}ller-Groeling}}, \ and\ \bibinfo {author} {\bibfnamefont {Hans~A}\
  \bibnamefont {Weidenm{\"u}ller}},\ }\bibfield  {title} {\enquote {\bibinfo
  {title} {Random-matrix theories in quantum physics: common concepts},}\
  }\href {\doibase 10.1016/S0370-1573(97)00088-4} {\bibfield  {journal}
  {\bibinfo  {journal} {Physics Reports}\ }\textbf {\bibinfo {volume} {299}},\
  \bibinfo {pages} {189} (\bibinfo {year} {1998})}\BibitemShut {NoStop}%
\bibitem [{\citenamefont {Kramer}\ and\ \citenamefont
  {MacKinnon}(1993)}]{kramer1993localization}%
  \BibitemOpen
  \bibfield  {author} {\bibinfo {author} {\bibfnamefont {Bernhard}\
  \bibnamefont {Kramer}}\ and\ \bibinfo {author} {\bibfnamefont {Angus}\
  \bibnamefont {MacKinnon}},\ }\bibfield  {title} {\enquote {\bibinfo {title}
  {Localization: theory and experiment},}\ }\href {\doibase
  10.1088/0034-4885/56/12/001} {\bibfield  {journal} {\bibinfo  {journal}
  {Reports on Progress in Physics}\ }\textbf {\bibinfo {volume} {56}},\
  \bibinfo {pages} {1469} (\bibinfo {year} {1993})}\BibitemShut {NoStop}%
\bibitem [{\citenamefont {Aubry}\ and\ \citenamefont
  {Andr{\'{e}}}(1980)}]{Aubry1980}%
  \BibitemOpen
  \bibfield  {author} {\bibinfo {author} {\bibfnamefont {Serge}\ \bibnamefont
  {Aubry}}\ and\ \bibinfo {author} {\bibfnamefont {Gilles}\ \bibnamefont
  {Andr{\'{e}}}},\ }\href@noop {} {\enquote {\bibinfo {title} {{Analyticity
  breaking and Anderson localization in incommensurate lattices}},}\ }
  (\bibinfo {year} {1980})\BibitemShut {NoStop}%
\bibitem [{\citenamefont {Harper}(1955)}]{Harper1955}%
  \BibitemOpen
  \bibfield  {author} {\bibinfo {author} {\bibfnamefont {P.~G.}\ \bibnamefont
  {Harper}},\ }\bibfield  {title} {\enquote {\bibinfo {title} {{Single band
  motion of conduction electrons in a uniform magnetic field}},}\ }\href
  {\doibase 10.1088/0370-1298/68/10/304} {\bibfield  {journal} {\bibinfo
  {journal} {Proceedings of the Physical Society. Section A}\ }\textbf
  {\bibinfo {volume} {68}},\ \bibinfo {pages} {874} (\bibinfo {year}
  {1955})}\BibitemShut {NoStop}%
\bibitem [{\citenamefont {Madsen}\ \emph {et~al.}(2013)\citenamefont {Madsen},
  \citenamefont {Bergholtz},\ and\ \citenamefont {Brouwer}}]{Brouwer2013}%
  \BibitemOpen
  \bibfield  {author} {\bibinfo {author} {\bibfnamefont {Kevin~A.}\
  \bibnamefont {Madsen}}, \bibinfo {author} {\bibfnamefont {Emil~J.}\
  \bibnamefont {Bergholtz}}, \ and\ \bibinfo {author} {\bibfnamefont {Piet~W.}\
  \bibnamefont {Brouwer}},\ }\bibfield  {title} {\enquote {\bibinfo {title}
  {{Topological equivalence of crystal and quasicrystal band structures}},}\
  }\href {\doibase 10.1103/PhysRevB.88.125118} {\bibfield  {journal} {\bibinfo
  {journal} {Phys. Rev. B}\ }\textbf {\bibinfo {volume} {88}},\ \bibinfo
  {pages} {125118} (\bibinfo {year} {2013})}\BibitemShut {NoStop}%
\bibitem [{\citenamefont {Lang}\ \emph {et~al.}(2012)\citenamefont {Lang},
  \citenamefont {Cai},\ and\ \citenamefont {Chen}}]{Chen2012}%
  \BibitemOpen
  \bibfield  {author} {\bibinfo {author} {\bibfnamefont {Li-Jun}\ \bibnamefont
  {Lang}}, \bibinfo {author} {\bibfnamefont {Xiaoming}\ \bibnamefont {Cai}}, \
  and\ \bibinfo {author} {\bibfnamefont {Shu}\ \bibnamefont {Chen}},\
  }\bibfield  {title} {\enquote {\bibinfo {title} {{Edge States and Topological
  Phases in One-Dimensional Optical Superlattices}},}\ }\href {\doibase
  10.1103/PhysRevLett.108.220401} {\bibfield  {journal} {\bibinfo  {journal}
  {Phys. Rev. Lett.}\ }\textbf {\bibinfo {volume} {108}},\ \bibinfo {pages}
  {220401} (\bibinfo {year} {2012})}\BibitemShut {NoStop}%
\bibitem [{\citenamefont {Vidal}\ \emph {et~al.}(1999)\citenamefont {Vidal},
  \citenamefont {Mouhanna},\ and\ \citenamefont {Giamarchi}}]{Giamarchi1999}%
  \BibitemOpen
  \bibfield  {author} {\bibinfo {author} {\bibfnamefont {Julien}\ \bibnamefont
  {Vidal}}, \bibinfo {author} {\bibfnamefont {Dominique}\ \bibnamefont
  {Mouhanna}}, \ and\ \bibinfo {author} {\bibfnamefont {Thierry}\ \bibnamefont
  {Giamarchi}},\ }\bibfield  {title} {\enquote {\bibinfo {title} {{Correlated
  Fermions in a One-Dimensional Quasiperiodic Potential}},}\ }\href {\doibase
  10.1103/PhysRevLett.83.3908} {\bibfield  {journal} {\bibinfo  {journal}
  {Phys. Rev. Lett.}\ }\textbf {\bibinfo {volume} {83}},\ \bibinfo {pages}
  {3908} (\bibinfo {year} {1999})}\BibitemShut {NoStop}%
\bibitem [{\citenamefont {Kraus}\ and\ \citenamefont
  {Zilberberg}(2012)}]{Zilbergerg2012}%
  \BibitemOpen
  \bibfield  {author} {\bibinfo {author} {\bibfnamefont {Yaacov~E.}\
  \bibnamefont {Kraus}}\ and\ \bibinfo {author} {\bibfnamefont {Oded}\
  \bibnamefont {Zilberberg}},\ }\bibfield  {title} {\enquote {\bibinfo {title}
  {{Topological Equivalence between the Fibonacci Quasicrystal and the Harper
  Model}},}\ }\href {\doibase 10.1103/PhysRevLett.109.116404} {\bibfield
  {journal} {\bibinfo  {journal} {Phys. Rev. Lett.}\ }\textbf {\bibinfo
  {volume} {109}},\ \bibinfo {pages} {116404} (\bibinfo {year}
  {2012})}\BibitemShut {NoStop}%
\bibitem [{\citenamefont {Ganeshan}\ \emph {et~al.}(2013)\citenamefont
  {Ganeshan}, \citenamefont {Sun},\ and\ \citenamefont
  {Das~Sarma}}]{Ganeshan2013}%
  \BibitemOpen
  \bibfield  {author} {\bibinfo {author} {\bibfnamefont {Sriram}\ \bibnamefont
  {Ganeshan}}, \bibinfo {author} {\bibfnamefont {Kai}\ \bibnamefont {Sun}}, \
  and\ \bibinfo {author} {\bibfnamefont {S.}~\bibnamefont {Das~Sarma}},\
  }\bibfield  {title} {\enquote {\bibinfo {title} {{Topological Zero-Energy
  Modes in Gapless Commensurate Aubry-Andr\'e-Harper Models}},}\ }\href
  {\doibase 10.1103/PhysRevLett.110.180403} {\bibfield  {journal} {\bibinfo
  {journal} {Phys. Rev. Lett.}\ }\textbf {\bibinfo {volume} {110}},\ \bibinfo
  {pages} {180403} (\bibinfo {year} {2013})}\BibitemShut {NoStop}%
\bibitem [{\citenamefont {Liu}\ \emph {et~al.}(2015)\citenamefont {Liu},
  \citenamefont {Ghosh},\ and\ \citenamefont {Chong}}]{Liu2015}%
  \BibitemOpen
  \bibfield  {author} {\bibinfo {author} {\bibfnamefont {Fangli}\ \bibnamefont
  {Liu}}, \bibinfo {author} {\bibfnamefont {Somnath}\ \bibnamefont {Ghosh}}, \
  and\ \bibinfo {author} {\bibfnamefont {Y.~D.}\ \bibnamefont {Chong}},\
  }\bibfield  {title} {\enquote {\bibinfo {title} {{Localization and adiabatic
  pumping in a generalized Aubry-Andr\'e-Harper model}},}\ }\href {\doibase
  10.1103/PhysRevB.91.014108} {\bibfield  {journal} {\bibinfo  {journal} {Phys.
  Rev. B}\ }\textbf {\bibinfo {volume} {91}},\ \bibinfo {pages} {014108}
  (\bibinfo {year} {2015})}\BibitemShut {NoStop}%
\bibitem [{\citenamefont {Zeng}\ \emph
  {et~al.}(2020{\natexlab{a}})\citenamefont {Zeng}, \citenamefont {Yang},\ and\
  \citenamefont {L\"u}}]{zeng2020-1}%
  \BibitemOpen
  \bibfield  {author} {\bibinfo {author} {\bibfnamefont {Qi-Bo}\ \bibnamefont
  {Zeng}}, \bibinfo {author} {\bibfnamefont {Yan-Bin}\ \bibnamefont {Yang}}, \
  and\ \bibinfo {author} {\bibfnamefont {Rong}\ \bibnamefont {L\"u}},\
  }\bibfield  {title} {\enquote {\bibinfo {title} {{Topological phases in
  one-dimensional nonreciprocal superlattices}},}\ }\href {\doibase
  10.1103/PhysRevB.101.125418} {\bibfield  {journal} {\bibinfo  {journal}
  {Phys. Rev. B}\ }\textbf {\bibinfo {volume} {101}},\ \bibinfo {pages}
  {125418} (\bibinfo {year} {2020}{\natexlab{a}})}\BibitemShut {NoStop}%
\bibitem [{\citenamefont {Lahini}\ \emph {et~al.}(2009)\citenamefont {Lahini},
  \citenamefont {Pugatch}, \citenamefont {Pozzi}, \citenamefont {Sorel},
  \citenamefont {Morandotti}, \citenamefont {Davidson},\ and\ \citenamefont
  {Silberberg}}]{Lahini2009}%
  \BibitemOpen
  \bibfield  {author} {\bibinfo {author} {\bibfnamefont {Y.}~\bibnamefont
  {Lahini}}, \bibinfo {author} {\bibfnamefont {R.}~\bibnamefont {Pugatch}},
  \bibinfo {author} {\bibfnamefont {F.}~\bibnamefont {Pozzi}}, \bibinfo
  {author} {\bibfnamefont {M.}~\bibnamefont {Sorel}}, \bibinfo {author}
  {\bibfnamefont {R.}~\bibnamefont {Morandotti}}, \bibinfo {author}
  {\bibfnamefont {N.}~\bibnamefont {Davidson}}, \ and\ \bibinfo {author}
  {\bibfnamefont {Y.}~\bibnamefont {Silberberg}},\ }\bibfield  {title}
  {\enquote {\bibinfo {title} {{Observation of a Localization Transition in
  Quasiperiodic Photonic Lattices}},}\ }\href {\doibase
  10.1103/PhysRevLett.103.013901} {\bibfield  {journal} {\bibinfo  {journal}
  {Phys. Rev. Lett.}\ }\textbf {\bibinfo {volume} {103}},\ \bibinfo {pages}
  {013901} (\bibinfo {year} {2009})}\BibitemShut {NoStop}%
\bibitem [{\citenamefont {Verbin}\ \emph {et~al.}(2013)\citenamefont {Verbin},
  \citenamefont {Zilberberg}, \citenamefont {Kraus}, \citenamefont {Lahini},\
  and\ \citenamefont {Silberberg}}]{Silberberg2013}%
  \BibitemOpen
  \bibfield  {author} {\bibinfo {author} {\bibfnamefont {Mor}\ \bibnamefont
  {Verbin}}, \bibinfo {author} {\bibfnamefont {Oded}\ \bibnamefont
  {Zilberberg}}, \bibinfo {author} {\bibfnamefont {Yaacov~E.}\ \bibnamefont
  {Kraus}}, \bibinfo {author} {\bibfnamefont {Yoav}\ \bibnamefont {Lahini}}, \
  and\ \bibinfo {author} {\bibfnamefont {Yaron}\ \bibnamefont {Silberberg}},\
  }\bibfield  {title} {\enquote {\bibinfo {title} {{Observation of Topological
  Phase Transitions in Photonic Quasicrystals}},}\ }\href {\doibase
  10.1103/PhysRevLett.110.076403} {\bibfield  {journal} {\bibinfo  {journal}
  {Phys. Rev. Lett.}\ }\textbf {\bibinfo {volume} {110}},\ \bibinfo {pages}
  {076403} (\bibinfo {year} {2013})}\BibitemShut {NoStop}%
\bibitem [{\citenamefont {Kraus}\ \emph {et~al.}(2012)\citenamefont {Kraus},
  \citenamefont {Lahini}, \citenamefont {Ringel}, \citenamefont {Verbin},\ and\
  \citenamefont {Zilberberg}}]{Kraus2012}%
  \BibitemOpen
  \bibfield  {author} {\bibinfo {author} {\bibfnamefont {Yaacov~E.}\
  \bibnamefont {Kraus}}, \bibinfo {author} {\bibfnamefont {Yoav}\ \bibnamefont
  {Lahini}}, \bibinfo {author} {\bibfnamefont {Zohar}\ \bibnamefont {Ringel}},
  \bibinfo {author} {\bibfnamefont {Mor}\ \bibnamefont {Verbin}}, \ and\
  \bibinfo {author} {\bibfnamefont {Oded}\ \bibnamefont {Zilberberg}},\
  }\bibfield  {title} {\enquote {\bibinfo {title} {{Topological States and
  Adiabatic Pumping in Quasicrystals}},}\ }\href {\doibase
  10.1103/PhysRevLett.109.106402} {\bibfield  {journal} {\bibinfo  {journal}
  {Phys. Rev. Lett.}\ }\textbf {\bibinfo {volume} {109}},\ \bibinfo {pages}
  {106402} (\bibinfo {year} {2012})}\BibitemShut {NoStop}%
\bibitem [{\citenamefont {Dal~Negro}\ \emph {et~al.}(2003)\citenamefont
  {Dal~Negro}, \citenamefont {Oton}, \citenamefont {Gaburro}, \citenamefont
  {Pavesi}, \citenamefont {Johnson}, \citenamefont {Lagendijk}, \citenamefont
  {Righini}, \citenamefont {Colocci},\ and\ \citenamefont {Wiersma}}]{Dal2003}%
  \BibitemOpen
  \bibfield  {author} {\bibinfo {author} {\bibfnamefont {Luca}\ \bibnamefont
  {Dal~Negro}}, \bibinfo {author} {\bibfnamefont {Claudio~J.}\ \bibnamefont
  {Oton}}, \bibinfo {author} {\bibfnamefont {Zeno}\ \bibnamefont {Gaburro}},
  \bibinfo {author} {\bibfnamefont {Lorenzo}\ \bibnamefont {Pavesi}}, \bibinfo
  {author} {\bibfnamefont {Patrick}\ \bibnamefont {Johnson}}, \bibinfo {author}
  {\bibfnamefont {Ad}~\bibnamefont {Lagendijk}}, \bibinfo {author}
  {\bibfnamefont {Roberto}\ \bibnamefont {Righini}}, \bibinfo {author}
  {\bibfnamefont {Marcello}\ \bibnamefont {Colocci}}, \ and\ \bibinfo {author}
  {\bibfnamefont {Diederik~S.}\ \bibnamefont {Wiersma}},\ }\bibfield  {title}
  {\enquote {\bibinfo {title} {{Light Transport through the Band-Edge States of
  Fibonacci Quasicrystals}},}\ }\href {\doibase 10.1103/PhysRevLett.90.055501}
  {\bibfield  {journal} {\bibinfo  {journal} {Phys. Rev. Lett.}\ }\textbf
  {\bibinfo {volume} {90}},\ \bibinfo {pages} {055501} (\bibinfo {year}
  {2003})}\BibitemShut {NoStop}%
\bibitem [{\citenamefont {Xiao}\ \emph {et~al.}(2021)\citenamefont {Xiao},
  \citenamefont {Xie}, \citenamefont {Dong}, \citenamefont {Chen},
  \citenamefont {Yi},\ and\ \citenamefont {Yan}}]{xiao2021}%
  \BibitemOpen
  \bibfield  {author} {\bibinfo {author} {\bibfnamefont {Teng}\ \bibnamefont
  {Xiao}}, \bibinfo {author} {\bibfnamefont {Dizhou}\ \bibnamefont {Xie}},
  \bibinfo {author} {\bibfnamefont {Zhaoli}\ \bibnamefont {Dong}}, \bibinfo
  {author} {\bibfnamefont {Tao}\ \bibnamefont {Chen}}, \bibinfo {author}
  {\bibfnamefont {Wei}\ \bibnamefont {Yi}}, \ and\ \bibinfo {author}
  {\bibfnamefont {Bo}~\bibnamefont {Yan}},\ }\bibfield  {title} {\enquote
  {\bibinfo {title} {{Observation of topological phase with critical
  localization in a quasi-periodic lattice}},}\ }\href {\doibase
  https://doi.org/10.1016/j.scib.2021.07.025} {\bibfield  {journal} {\bibinfo
  {journal} {Science Bulletin}\ }\textbf {\bibinfo {volume} {66}},\ \bibinfo
  {pages} {2175} (\bibinfo {year} {2021})}\BibitemShut {NoStop}%
\bibitem [{\citenamefont {Roati}\ \emph {et~al.}(2008)\citenamefont {Roati},
  \citenamefont {D'Errico}, \citenamefont {Fallani}, \citenamefont {Fattori},
  \citenamefont {Fort}, \citenamefont {Zaccanti}, \citenamefont {Modugno},
  \citenamefont {Modugno},\ and\ \citenamefont {Inguscio}}]{Roati2008}%
  \BibitemOpen
  \bibfield  {author} {\bibinfo {author} {\bibfnamefont {Giacomo}\ \bibnamefont
  {Roati}}, \bibinfo {author} {\bibfnamefont {Chiara}\ \bibnamefont
  {D'Errico}}, \bibinfo {author} {\bibfnamefont {Leonardo}\ \bibnamefont
  {Fallani}}, \bibinfo {author} {\bibfnamefont {Marco}\ \bibnamefont
  {Fattori}}, \bibinfo {author} {\bibfnamefont {Chiara}\ \bibnamefont {Fort}},
  \bibinfo {author} {\bibfnamefont {Matteo}\ \bibnamefont {Zaccanti}}, \bibinfo
  {author} {\bibfnamefont {Giovanni}\ \bibnamefont {Modugno}}, \bibinfo
  {author} {\bibfnamefont {Michele}\ \bibnamefont {Modugno}}, \ and\ \bibinfo
  {author} {\bibfnamefont {Massimo}\ \bibnamefont {Inguscio}},\ }\bibfield
  {title} {\enquote {\bibinfo {title} {{Anderson localization of a
  non-interacting Bose-Einstein condensate}},}\ }\href {\doibase
  10.1038/nature07071} {\bibfield  {journal} {\bibinfo  {journal} {Nature}\
  }\textbf {\bibinfo {volume} {453}},\ \bibinfo {pages} {895} (\bibinfo {year}
  {2008})}\BibitemShut {NoStop}%
\bibitem [{\citenamefont {Modugno}(2010)}]{Modugno2010}%
  \BibitemOpen
  \bibfield  {author} {\bibinfo {author} {\bibfnamefont {Giovanni}\
  \bibnamefont {Modugno}},\ }\bibfield  {title} {\enquote {\bibinfo {title}
  {{Anderson localization in Bose{\textendash}Einstein condensates}},}\ }\href
  {\doibase 10.1088/0034-4885/73/10/102401} {\bibfield  {journal} {\bibinfo
  {journal} {Reports on Progress in Physics}\ }\textbf {\bibinfo {volume}
  {73}},\ \bibinfo {pages} {102401} (\bibinfo {year} {2010})}\BibitemShut
  {NoStop}%
\bibitem [{\citenamefont {Biddle}\ and\ \citenamefont
  {Das~Sarma}(2010)}]{Biddle2010}%
  \BibitemOpen
  \bibfield  {author} {\bibinfo {author} {\bibfnamefont {J.}~\bibnamefont
  {Biddle}}\ and\ \bibinfo {author} {\bibfnamefont {S.}~\bibnamefont
  {Das~Sarma}},\ }\bibfield  {title} {\enquote {\bibinfo {title} {{Predicted
  Mobility Edges in One-Dimensional Incommensurate Optical Lattices: An Exactly
  Solvable Model of Anderson Localization}},}\ }\href {\doibase
  10.1103/PhysRevLett.104.070601} {\bibfield  {journal} {\bibinfo  {journal}
  {Phys. Rev. Lett.}\ }\textbf {\bibinfo {volume} {104}},\ \bibinfo {pages}
  {070601} (\bibinfo {year} {2010})}\BibitemShut {NoStop}%
\bibitem [{\citenamefont {Liu}\ \emph {et~al.}(2020{\natexlab{a}})\citenamefont
  {Liu}, \citenamefont {Guo}, \citenamefont {Pu},\ and\ \citenamefont
  {Longhi}}]{Longhi2020}%
  \BibitemOpen
  \bibfield  {author} {\bibinfo {author} {\bibfnamefont {Tong}\ \bibnamefont
  {Liu}}, \bibinfo {author} {\bibfnamefont {Hao}\ \bibnamefont {Guo}}, \bibinfo
  {author} {\bibfnamefont {Yong}\ \bibnamefont {Pu}}, \ and\ \bibinfo {author}
  {\bibfnamefont {Stefano}\ \bibnamefont {Longhi}},\ }\bibfield  {title}
  {\enquote {\bibinfo {title} {{Generalized Aubry-Andr\'e self-duality and
  mobility edges in non-Hermitian quasiperiodic lattices}},}\ }\href {\doibase
  10.1103/PhysRevB.102.024205} {\bibfield  {journal} {\bibinfo  {journal}
  {Phys. Rev. B}\ }\textbf {\bibinfo {volume} {102}},\ \bibinfo {pages}
  {024205} (\bibinfo {year} {2020}{\natexlab{a}})}\BibitemShut {NoStop}%
\bibitem [{\citenamefont {Cestari}\ \emph {et~al.}(2016)\citenamefont
  {Cestari}, \citenamefont {Foerster},\ and\ \citenamefont
  {Gusm\~ao}}]{Cestari2016}%
  \BibitemOpen
  \bibfield  {author} {\bibinfo {author} {\bibfnamefont {J.~C.~C.}\
  \bibnamefont {Cestari}}, \bibinfo {author} {\bibfnamefont {A.}~\bibnamefont
  {Foerster}}, \ and\ \bibinfo {author} {\bibfnamefont {M.~A.}\ \bibnamefont
  {Gusm\~ao}},\ }\bibfield  {title} {\enquote {\bibinfo {title} {{Fate of
  topological states in incommensurate generalized Aubry-Andr\'e models}},}\
  }\href {\doibase 10.1103/PhysRevB.93.205441} {\bibfield  {journal} {\bibinfo
  {journal} {Phys. Rev. B}\ }\textbf {\bibinfo {volume} {93}},\ \bibinfo
  {pages} {205441} (\bibinfo {year} {2016})}\BibitemShut {NoStop}%
\bibitem [{\citenamefont {Liu}\ \emph {et~al.}(2016)\citenamefont {Liu},
  \citenamefont {Wang},\ and\ \citenamefont {Xianlong}}]{Liutong2016}%
  \BibitemOpen
  \bibfield  {author} {\bibinfo {author} {\bibfnamefont {Tong}\ \bibnamefont
  {Liu}}, \bibinfo {author} {\bibfnamefont {Pei}\ \bibnamefont {Wang}}, \ and\
  \bibinfo {author} {\bibfnamefont {Gao}\ \bibnamefont {Xianlong}},\
  }\href@noop {} {\enquote {\bibinfo {title} {{Phase diagram of the
  off-diagonal Aubry-Andr\'e model}},}\ } (\bibinfo {year} {2016}),\ \Eprint
  {http://arxiv.org/abs/1609.06939} {arXiv:1609.06939 [cond-mat.dis-nn]}
  \BibitemShut {NoStop}%
\bibitem [{\citenamefont {Hatano}\ and\ \citenamefont
  {Nelson}(1996)}]{Hatano1996}%
  \BibitemOpen
  \bibfield  {author} {\bibinfo {author} {\bibfnamefont {Naomichi}\
  \bibnamefont {Hatano}}\ and\ \bibinfo {author} {\bibfnamefont {David~R.}\
  \bibnamefont {Nelson}},\ }\bibfield  {title} {\enquote {\bibinfo {title}
  {{Localization Transitions in Non-Hermitian Quantum Mechanics}},}\ }\href
  {\doibase 10.1103/PhysRevLett.77.570} {\bibfield  {journal} {\bibinfo
  {journal} {Phys. Rev. Lett.}\ }\textbf {\bibinfo {volume} {77}},\ \bibinfo
  {pages} {570} (\bibinfo {year} {1996})}\BibitemShut {NoStop}%
\bibitem [{\citenamefont {Longhi}(2021{\natexlab{a}})}]{Longhi2021}%
  \BibitemOpen
  \bibfield  {author} {\bibinfo {author} {\bibfnamefont {Stefano}\ \bibnamefont
  {Longhi}},\ }\bibfield  {title} {\enquote {\bibinfo {title} {{Non-Hermitian
  skin effect beyond the tight-binding models}},}\ }\href {\doibase
  10.1103/PhysRevB.104.125109} {\bibfield  {journal} {\bibinfo  {journal}
  {Phys. Rev. B}\ }\textbf {\bibinfo {volume} {104}},\ \bibinfo {pages}
  {125109} (\bibinfo {year} {2021}{\natexlab{a}})}\BibitemShut {NoStop}%
\bibitem [{\citenamefont {Longhi}(2021{\natexlab{b}})}]{Longhi2021-1}%
  \BibitemOpen
  \bibfield  {author} {\bibinfo {author} {\bibfnamefont {Stefano}\ \bibnamefont
  {Longhi}},\ }\bibfield  {title} {\enquote {\bibinfo {title} {{Spectral
  deformations in non-Hermitian lattices with disorder and skin effect: A
  solvable model}},}\ }\href {\doibase 10.1103/PhysRevB.103.144202} {\bibfield
  {journal} {\bibinfo  {journal} {Phys. Rev. B}\ }\textbf {\bibinfo {volume}
  {103}},\ \bibinfo {pages} {144202} (\bibinfo {year}
  {2021}{\natexlab{b}})}\BibitemShut {NoStop}%
\bibitem [{\citenamefont {Longhi}(2021{\natexlab{c}})}]{Longhi2021-2}%
  \BibitemOpen
  \bibfield  {author} {\bibinfo {author} {\bibfnamefont {Stefano}\ \bibnamefont
  {Longhi}},\ }\bibfield  {title} {\enquote {\bibinfo {title} {{Phase
  transitions in a non-Hermitian Aubry-Andr\'e-Harper model}},}\ }\href
  {\doibase 10.1103/PhysRevB.103.054203} {\bibfield  {journal} {\bibinfo
  {journal} {Phys. Rev. B}\ }\textbf {\bibinfo {volume} {103}},\ \bibinfo
  {pages} {054203} (\bibinfo {year} {2021}{\natexlab{c}})}\BibitemShut
  {NoStop}%
\bibitem [{\citenamefont {Longhi}(2019{\natexlab{a}})}]{Longhi2019}%
  \BibitemOpen
  \bibfield  {author} {\bibinfo {author} {\bibfnamefont {Stefano}\ \bibnamefont
  {Longhi}},\ }\bibfield  {title} {\enquote {\bibinfo {title} {{Metal-insulator
  phase transition in a non-Hermitian Aubry-Andr\'e-Harper model}},}\ }\href
  {\doibase 10.1103/PhysRevB.100.125157} {\bibfield  {journal} {\bibinfo
  {journal} {Phys. Rev. B}\ }\textbf {\bibinfo {volume} {100}},\ \bibinfo
  {pages} {125157} (\bibinfo {year} {2019}{\natexlab{a}})}\BibitemShut
  {NoStop}%
\bibitem [{\citenamefont {Longhi}(2019{\natexlab{b}})}]{Longhi2019-1}%
  \BibitemOpen
  \bibfield  {author} {\bibinfo {author} {\bibfnamefont {S.}~\bibnamefont
  {Longhi}},\ }\bibfield  {title} {\enquote {\bibinfo {title} {{Topological
  Phase Transition in non-Hermitian Quasicrystals}},}\ }\href {\doibase
  10.1103/PhysRevLett.122.237601} {\bibfield  {journal} {\bibinfo  {journal}
  {Phys. Rev. Lett.}\ }\textbf {\bibinfo {volume} {122}},\ \bibinfo {pages}
  {237601} (\bibinfo {year} {2019}{\natexlab{b}})}\BibitemShut {NoStop}%
\bibitem [{\citenamefont {Bender}\ and\ \citenamefont
  {Boettcher}(1998)}]{Bender1998}%
  \BibitemOpen
  \bibfield  {author} {\bibinfo {author} {\bibfnamefont {Carl~M.}\ \bibnamefont
  {Bender}}\ and\ \bibinfo {author} {\bibfnamefont {Stefan}\ \bibnamefont
  {Boettcher}},\ }\bibfield  {title} {\enquote {\bibinfo {title} {{Real Spectra
  in Non-Hermitian Hamiltonians Having ${P}{T}$ Symmetry}},}\ }\href {\doibase
  10.1103/PhysRevLett.80.5243} {\bibfield  {journal} {\bibinfo  {journal}
  {Phys. Rev. Lett.}\ }\textbf {\bibinfo {volume} {80}},\ \bibinfo {pages}
  {5243} (\bibinfo {year} {1998})}\BibitemShut {NoStop}%
\bibitem [{\citenamefont {Zeng}\ \emph {et~al.}(2017)\citenamefont {Zeng},
  \citenamefont {Chen},\ and\ \citenamefont {L\"u}}]{Chen2017}%
  \BibitemOpen
  \bibfield  {author} {\bibinfo {author} {\bibfnamefont {Qi-Bo}\ \bibnamefont
  {Zeng}}, \bibinfo {author} {\bibfnamefont {Shu}\ \bibnamefont {Chen}}, \ and\
  \bibinfo {author} {\bibfnamefont {Rong}\ \bibnamefont {L\"u}},\ }\bibfield
  {title} {\enquote {\bibinfo {title} {{Anderson localization in the
  non-Hermitian Aubry-Andr\'e-Harper model with physical gain and loss}},}\
  }\href {\doibase 10.1103/PhysRevA.95.062118} {\bibfield  {journal} {\bibinfo
  {journal} {Phys. Rev. A}\ }\textbf {\bibinfo {volume} {95}},\ \bibinfo
  {pages} {062118} (\bibinfo {year} {2017})}\BibitemShut {NoStop}%
\bibitem [{\citenamefont {Xu}\ \emph {et~al.}(2017)\citenamefont {Xu},
  \citenamefont {Zhang},\ and\ \citenamefont {Chen}}]{Chen2017-2}%
  \BibitemOpen
  \bibfield  {author} {\bibinfo {author} {\bibfnamefont {Zhihao}\ \bibnamefont
  {Xu}}, \bibinfo {author} {\bibfnamefont {Yunbo}\ \bibnamefont {Zhang}}, \
  and\ \bibinfo {author} {\bibfnamefont {Shu}\ \bibnamefont {Chen}},\
  }\bibfield  {title} {\enquote {\bibinfo {title} {{Topological phase
  transition and charge pumping in a one-dimensional periodically driven
  optical lattice}},}\ }\href {\doibase 10.1103/PhysRevA.96.013606} {\bibfield
  {journal} {\bibinfo  {journal} {Phys. Rev. A}\ }\textbf {\bibinfo {volume}
  {96}},\ \bibinfo {pages} {013606} (\bibinfo {year} {2017})}\BibitemShut
  {NoStop}%
\bibitem [{\citenamefont {Liu}\ \emph {et~al.}(2020{\natexlab{b}})\citenamefont
  {Liu}, \citenamefont {Jiang}, \citenamefont {Cao},\ and\ \citenamefont
  {Chen}}]{Chen2020}%
  \BibitemOpen
  \bibfield  {author} {\bibinfo {author} {\bibfnamefont {Yanxia}\ \bibnamefont
  {Liu}}, \bibinfo {author} {\bibfnamefont {Xiang-Ping}\ \bibnamefont {Jiang}},
  \bibinfo {author} {\bibfnamefont {Junpeng}\ \bibnamefont {Cao}}, \ and\
  \bibinfo {author} {\bibfnamefont {Shu}\ \bibnamefont {Chen}},\ }\bibfield
  {title} {\enquote {\bibinfo {title} {{Non-Hermitian mobility edges in
  one-dimensional quasicrystals with parity-time symmetry}},}\ }\href {\doibase
  10.1103/PhysRevB.101.174205} {\bibfield  {journal} {\bibinfo  {journal}
  {Phys. Rev. B}\ }\textbf {\bibinfo {volume} {101}},\ \bibinfo {pages}
  {174205} (\bibinfo {year} {2020}{\natexlab{b}})}\BibitemShut {NoStop}%
\bibitem [{\citenamefont {Guo}\ \emph {et~al.}(2021)\citenamefont {Guo},
  \citenamefont {Liu}, \citenamefont {Zhao}, \citenamefont {Liu},\ and\
  \citenamefont {Chen}}]{Chen2021}%
  \BibitemOpen
  \bibfield  {author} {\bibinfo {author} {\bibfnamefont {Cui-Xian}\
  \bibnamefont {Guo}}, \bibinfo {author} {\bibfnamefont {Chun-Hui}\
  \bibnamefont {Liu}}, \bibinfo {author} {\bibfnamefont {Xiao-Ming}\
  \bibnamefont {Zhao}}, \bibinfo {author} {\bibfnamefont {Yanxia}\ \bibnamefont
  {Liu}}, \ and\ \bibinfo {author} {\bibfnamefont {Shu}\ \bibnamefont {Chen}},\
  }\bibfield  {title} {\enquote {\bibinfo {title} {{Exact Solution of
  Non-Hermitian Systems with Generalized Boundary Conditions: Size-Dependent
  Boundary Effect and Fragility of the Skin Effect}},}\ }\href {\doibase
  10.1103/PhysRevLett.127.116801} {\bibfield  {journal} {\bibinfo  {journal}
  {Phys. Rev. Lett.}\ }\textbf {\bibinfo {volume} {127}},\ \bibinfo {pages}
  {116801} (\bibinfo {year} {2021})}\BibitemShut {NoStop}%
\bibitem [{\citenamefont {Liu}\ \emph {et~al.}(2021{\natexlab{a}})\citenamefont
  {Liu}, \citenamefont {Zeng}, \citenamefont {Li},\ and\ \citenamefont
  {Chen}}]{Chen2021-1}%
  \BibitemOpen
  \bibfield  {author} {\bibinfo {author} {\bibfnamefont {Yanxia}\ \bibnamefont
  {Liu}}, \bibinfo {author} {\bibfnamefont {Yumeng}\ \bibnamefont {Zeng}},
  \bibinfo {author} {\bibfnamefont {Linhu}\ \bibnamefont {Li}}, \ and\ \bibinfo
  {author} {\bibfnamefont {Shu}\ \bibnamefont {Chen}},\ }\bibfield  {title}
  {\enquote {\bibinfo {title} {{Exact solution of the single impurity problem
  in nonreciprocal lattices: Impurity-induced size-dependent non-Hermitian skin
  effect}},}\ }\href {\doibase 10.1103/PhysRevB.104.085401} {\bibfield
  {journal} {\bibinfo  {journal} {Phys. Rev. B}\ }\textbf {\bibinfo {volume}
  {104}},\ \bibinfo {pages} {085401} (\bibinfo {year}
  {2021}{\natexlab{a}})}\BibitemShut {NoStop}%
\bibitem [{\citenamefont {Liu}\ \emph {et~al.}(2021{\natexlab{b}})\citenamefont
  {Liu}, \citenamefont {Wang}, \citenamefont {Zheng},\ and\ \citenamefont
  {Chen}}]{Chen2021-2}%
  \BibitemOpen
  \bibfield  {author} {\bibinfo {author} {\bibfnamefont {Yanxia}\ \bibnamefont
  {Liu}}, \bibinfo {author} {\bibfnamefont {Yongjian}\ \bibnamefont {Wang}},
  \bibinfo {author} {\bibfnamefont {Zuohuan}\ \bibnamefont {Zheng}}, \ and\
  \bibinfo {author} {\bibfnamefont {Shu}\ \bibnamefont {Chen}},\ }\bibfield
  {title} {\enquote {\bibinfo {title} {{Exact non-Hermitian mobility edges in
  one-dimensional quasicrystal lattice with exponentially decaying hopping and
  its dual lattice}},}\ }\href {\doibase 10.1103/PhysRevB.103.134208}
  {\bibfield  {journal} {\bibinfo  {journal} {Phys. Rev. B}\ }\textbf {\bibinfo
  {volume} {103}},\ \bibinfo {pages} {134208} (\bibinfo {year}
  {2021}{\natexlab{b}})}\BibitemShut {NoStop}%
\bibitem [{\citenamefont {Liu}\ \emph {et~al.}(2021{\natexlab{c}})\citenamefont
  {Liu}, \citenamefont {Wang}, \citenamefont {Liu}, \citenamefont {Zhou},\ and\
  \citenamefont {Chen}}]{Chen2021-3}%
  \BibitemOpen
  \bibfield  {author} {\bibinfo {author} {\bibfnamefont {Yanxia}\ \bibnamefont
  {Liu}}, \bibinfo {author} {\bibfnamefont {Yucheng}\ \bibnamefont {Wang}},
  \bibinfo {author} {\bibfnamefont {Xiong-Jun}\ \bibnamefont {Liu}}, \bibinfo
  {author} {\bibfnamefont {Qi}~\bibnamefont {Zhou}}, \ and\ \bibinfo {author}
  {\bibfnamefont {Shu}\ \bibnamefont {Chen}},\ }\bibfield  {title} {\enquote
  {\bibinfo {title} {{Exact mobility edges, ${P}{T}$-symmetry breaking, and
  skin effect in one-dimensional non-Hermitian quasicrystals}},}\ }\href
  {\doibase 10.1103/PhysRevB.103.014203} {\bibfield  {journal} {\bibinfo
  {journal} {Phys. Rev. B}\ }\textbf {\bibinfo {volume} {103}},\ \bibinfo
  {pages} {014203} (\bibinfo {year} {2021}{\natexlab{c}})}\BibitemShut
  {NoStop}%
\bibitem [{\citenamefont {Zeng}\ \emph
  {et~al.}(2020{\natexlab{b}})\citenamefont {Zeng}, \citenamefont {Yang},\ and\
  \citenamefont {Xu}}]{zeng2020-2}%
  \BibitemOpen
  \bibfield  {author} {\bibinfo {author} {\bibfnamefont {Qi-Bo}\ \bibnamefont
  {Zeng}}, \bibinfo {author} {\bibfnamefont {Yan-Bin}\ \bibnamefont {Yang}}, \
  and\ \bibinfo {author} {\bibfnamefont {Yong}\ \bibnamefont {Xu}},\ }\bibfield
   {title} {\enquote {\bibinfo {title} {{Topological phases in non-Hermitian
  Aubry-Andr\'e-Harper models}},}\ }\href {\doibase
  10.1103/PhysRevB.101.020201} {\bibfield  {journal} {\bibinfo  {journal}
  {Phys. Rev. B}\ }\textbf {\bibinfo {volume} {101}},\ \bibinfo {pages}
  {020201} (\bibinfo {year} {2020}{\natexlab{b}})}\BibitemShut {NoStop}%
\bibitem [{\citenamefont {Bernardini}\ and\ \citenamefont
  {Bertolami}(2022)}]{PhysRevA.105.032207}%
  \BibitemOpen
  \bibfield  {author} {\bibinfo {author} {\bibfnamefont {A.~E.}\ \bibnamefont
  {Bernardini}}\ and\ \bibinfo {author} {\bibfnamefont {O.}~\bibnamefont
  {Bertolami}},\ }\bibfield  {title} {\enquote {\bibinfo {title} {Generalized
  phase-space description of nonlinear hamiltonian systems and harper-like
  dynamics},}\ }\href {\doibase 10.1103/PhysRevA.105.032207} {\bibfield
  {journal} {\bibinfo  {journal} {Phys. Rev. A}\ }\textbf {\bibinfo {volume}
  {105}},\ \bibinfo {pages} {032207} (\bibinfo {year} {2022})}\BibitemShut
  {NoStop}%
\bibitem [{\citenamefont {Hamazaki}\ \emph {et~al.}(2019)\citenamefont
  {Hamazaki}, \citenamefont {Kawabata},\ and\ \citenamefont {Ueda}}]{Ueda2019}%
  \BibitemOpen
  \bibfield  {author} {\bibinfo {author} {\bibfnamefont {Ryusuke}\ \bibnamefont
  {Hamazaki}}, \bibinfo {author} {\bibfnamefont {Kohei}\ \bibnamefont
  {Kawabata}}, \ and\ \bibinfo {author} {\bibfnamefont {Masahito}\ \bibnamefont
  {Ueda}},\ }\bibfield  {title} {\enquote {\bibinfo {title} {{Non-Hermitian
  Many-Body Localization}},}\ }\href {\doibase 10.1103/PhysRevLett.123.090603}
  {\bibfield  {journal} {\bibinfo  {journal} {Phys. Rev. Lett.}\ }\textbf
  {\bibinfo {volume} {123}},\ \bibinfo {pages} {090603} (\bibinfo {year}
  {2019})}\BibitemShut {NoStop}%
\bibitem [{\citenamefont {Tang}\ \emph {et~al.}(2021)\citenamefont {Tang},
  \citenamefont {Zhang}, \citenamefont {Zhang},\ and\ \citenamefont
  {Zhang}}]{Zhang2021}%
  \BibitemOpen
  \bibfield  {author} {\bibinfo {author} {\bibfnamefont {Ling-Zhi}\
  \bibnamefont {Tang}}, \bibinfo {author} {\bibfnamefont {Guo-Qing}\
  \bibnamefont {Zhang}}, \bibinfo {author} {\bibfnamefont {Ling-Feng}\
  \bibnamefont {Zhang}}, \ and\ \bibinfo {author} {\bibfnamefont {Dan-Wei}\
  \bibnamefont {Zhang}},\ }\bibfield  {title} {\enquote {\bibinfo {title}
  {{Localization and topological transitions in non-Hermitian quasiperiodic
  lattices}},}\ }\href {\doibase 10.1103/PhysRevA.103.033325} {\bibfield
  {journal} {\bibinfo  {journal} {Phys. Rev. A}\ }\textbf {\bibinfo {volume}
  {103}},\ \bibinfo {pages} {033325} (\bibinfo {year} {2021})}\BibitemShut
  {NoStop}%
\bibitem [{\citenamefont {Biddle}\ \emph {et~al.}(2011)\citenamefont {Biddle},
  \citenamefont {Priour}, \citenamefont {Wang},\ and\ \citenamefont
  {Das~Sarma}}]{Biddle2011}%
  \BibitemOpen
  \bibfield  {author} {\bibinfo {author} {\bibfnamefont {J.}~\bibnamefont
  {Biddle}}, \bibinfo {author} {\bibfnamefont {D.~J.}\ \bibnamefont {Priour}},
  \bibinfo {author} {\bibfnamefont {B.}~\bibnamefont {Wang}}, \ and\ \bibinfo
  {author} {\bibfnamefont {S.}~\bibnamefont {Das~Sarma}},\ }\bibfield  {title}
  {\enquote {\bibinfo {title} {{Localization in one-dimensional lattices with
  non-nearest-neighbor hopping: Generalized Anderson and Aubry-Andr\'e
  models}},}\ }\href {\doibase 10.1103/PhysRevB.83.075105} {\bibfield
  {journal} {\bibinfo  {journal} {Phys. Rev. B}\ }\textbf {\bibinfo {volume}
  {83}},\ \bibinfo {pages} {075105} (\bibinfo {year} {2011})}\BibitemShut
  {NoStop}%
\bibitem [{\citenamefont {Liu}\ \emph {et~al.}(2021{\natexlab{d}})\citenamefont
  {Liu}, \citenamefont {Cheng}, \citenamefont {Guo},\ and\ \citenamefont
  {Xianlong}}]{Liutong2021}%
  \BibitemOpen
  \bibfield  {author} {\bibinfo {author} {\bibfnamefont {Tong}\ \bibnamefont
  {Liu}}, \bibinfo {author} {\bibfnamefont {Shujie}\ \bibnamefont {Cheng}},
  \bibinfo {author} {\bibfnamefont {Hao}\ \bibnamefont {Guo}}, \ and\ \bibinfo
  {author} {\bibfnamefont {Gao}\ \bibnamefont {Xianlong}},\ }\bibfield  {title}
  {\enquote {\bibinfo {title} {{Fate of Majorana zero modes, exact location of
  critical states, and unconventional real-complex transition in non-Hermitian
  quasiperiodic lattices}},}\ }\href {\doibase 10.1103/PhysRevB.103.104203}
  {\bibfield  {journal} {\bibinfo  {journal} {Phys. Rev. B}\ }\textbf {\bibinfo
  {volume} {103}},\ \bibinfo {pages} {104203} (\bibinfo {year}
  {2021}{\natexlab{d}})}\BibitemShut {NoStop}%
\bibitem [{\citenamefont {Fyodorov}\ and\ \citenamefont
  {Mirlin}(1993)}]{PhysRevLett.71.412}%
  \BibitemOpen
  \bibfield  {author} {\bibinfo {author} {\bibfnamefont {Yan~V.}\ \bibnamefont
  {Fyodorov}}\ and\ \bibinfo {author} {\bibfnamefont {Alexander~D.}\
  \bibnamefont {Mirlin}},\ }\bibfield  {title} {\enquote {\bibinfo {title}
  {Level-to-level fluctuations of the inverse participation ratio in finite
  quasi 1d disordered systems},}\ }\href {\doibase 10.1103/PhysRevLett.71.412}
  {\bibfield  {journal} {\bibinfo  {journal} {Phys. Rev. Lett.}\ }\textbf
  {\bibinfo {volume} {71}},\ \bibinfo {pages} {412} (\bibinfo {year}
  {1993})}\BibitemShut {NoStop}%
\bibitem [{\citenamefont {Evers}\ and\ \citenamefont
  {Mirlin}(2000)}]{PhysRevLett.84.3690}%
  \BibitemOpen
  \bibfield  {author} {\bibinfo {author} {\bibfnamefont {F.}~\bibnamefont
  {Evers}}\ and\ \bibinfo {author} {\bibfnamefont {A.~D.}\ \bibnamefont
  {Mirlin}},\ }\bibfield  {title} {\enquote {\bibinfo {title} {Fluctuations of
  the inverse participation ratio at the anderson transition},}\ }\href
  {\doibase 10.1103/PhysRevLett.84.3690} {\bibfield  {journal} {\bibinfo
  {journal} {Phys. Rev. Lett.}\ }\textbf {\bibinfo {volume} {84}},\ \bibinfo
  {pages} {3690} (\bibinfo {year} {2000})}\BibitemShut {NoStop}%
\bibitem [{\citenamefont {Wang}\ \emph {et~al.}(2016)\citenamefont {Wang},
  \citenamefont {Liu}, \citenamefont {Xianlong},\ and\ \citenamefont
  {Hu}}]{Gao2016}%
  \BibitemOpen
  \bibfield  {author} {\bibinfo {author} {\bibfnamefont {Jun}\ \bibnamefont
  {Wang}}, \bibinfo {author} {\bibfnamefont {Xia-Ji}\ \bibnamefont {Liu}},
  \bibinfo {author} {\bibfnamefont {Gao}\ \bibnamefont {Xianlong}}, \ and\
  \bibinfo {author} {\bibfnamefont {Hui}\ \bibnamefont {Hu}},\ }\bibfield
  {title} {\enquote {\bibinfo {title} {{Phase diagram of a non-Abelian
  Aubry-Andr\'e-Harper model with $p$-wave superfluidity}},}\ }\href {\doibase
  10.1103/PhysRevB.93.104504} {\bibfield  {journal} {\bibinfo  {journal} {Phys.
  Rev. B}\ }\textbf {\bibinfo {volume} {93}},\ \bibinfo {pages} {104504}
  (\bibinfo {year} {2016})}\BibitemShut {NoStop}%
\bibitem [{\citenamefont {Li}\ and\ \citenamefont
  {Das~Sarma}(2020)}]{Lixiao2020}%
  \BibitemOpen
  \bibfield  {author} {\bibinfo {author} {\bibfnamefont {Xiao}\ \bibnamefont
  {Li}}\ and\ \bibinfo {author} {\bibfnamefont {S.}~\bibnamefont {Das~Sarma}},\
  }\bibfield  {title} {\enquote {\bibinfo {title} {{Mobility edge and
  intermediate phase in one-dimensional incommensurate lattice potentials}},}\
  }\href {\doibase 10.1103/PhysRevB.101.064203} {\bibfield  {journal} {\bibinfo
   {journal} {Phys. Rev. B}\ }\textbf {\bibinfo {volume} {101}},\ \bibinfo
  {pages} {064203} (\bibinfo {year} {2020})}\BibitemShut {NoStop}%
\bibitem [{\citenamefont {Weidemann}\ \emph {et~al.}(2022)\citenamefont
  {Weidemann}, \citenamefont {Kremer}, \citenamefont {Longhi},\ and\
  \citenamefont {Szameit}}]{weidemann2022topological}%
  \BibitemOpen
  \bibfield  {author} {\bibinfo {author} {\bibfnamefont {Sebastian}\
  \bibnamefont {Weidemann}}, \bibinfo {author} {\bibfnamefont {Mark}\
  \bibnamefont {Kremer}}, \bibinfo {author} {\bibfnamefont {Stefano}\
  \bibnamefont {Longhi}}, \ and\ \bibinfo {author} {\bibfnamefont {Alexander}\
  \bibnamefont {Szameit}},\ }\bibfield  {title} {\enquote {\bibinfo {title}
  {Topological triple phase transition in non-hermitian floquet
  quasicrystals},}\ }\href {\doibase 10.1038/s41586-021-04253-0} {\bibfield
  {journal} {\bibinfo  {journal} {Nature}\ }\textbf {\bibinfo {volume} {601}},\
  \bibinfo {pages} {354} (\bibinfo {year} {2022})}\BibitemShut {NoStop}%
\bibitem [{\citenamefont {Harter}\ \emph {et~al.}(2016)\citenamefont {Harter},
  \citenamefont {Lee},\ and\ \citenamefont {Joglekar}}]{Harter2016}%
  \BibitemOpen
  \bibfield  {author} {\bibinfo {author} {\bibfnamefont {Andrew~K.}\
  \bibnamefont {Harter}}, \bibinfo {author} {\bibfnamefont {Tony~E.}\
  \bibnamefont {Lee}}, \ and\ \bibinfo {author} {\bibfnamefont {Yogesh~N.}\
  \bibnamefont {Joglekar}},\ }\bibfield  {title} {\enquote {\bibinfo {title}
  {$\mathcal{PT}$-breaking threshold in spatially asymmetric aubry-andr\'e and
  harper models: Hidden symmetry and topological states},}\ }\href {\doibase
  10.1103/PhysRevA.93.062101} {\bibfield  {journal} {\bibinfo  {journal} {Phys.
  Rev. A}\ }\textbf {\bibinfo {volume} {93}},\ \bibinfo {pages} {062101}
  (\bibinfo {year} {2016})}\BibitemShut {NoStop}%
\bibitem [{\citenamefont {Schiffer}\ \emph {et~al.}(2021)\citenamefont
  {Schiffer}, \citenamefont {Liu}, \citenamefont {Hu},\ and\ \citenamefont
  {Wang}}]{Schiffer2021}%
  \BibitemOpen
  \bibfield  {author} {\bibinfo {author} {\bibfnamefont {Sebastian}\
  \bibnamefont {Schiffer}}, \bibinfo {author} {\bibfnamefont {Xia-Ji}\
  \bibnamefont {Liu}}, \bibinfo {author} {\bibfnamefont {Hui}\ \bibnamefont
  {Hu}}, \ and\ \bibinfo {author} {\bibfnamefont {Jia}\ \bibnamefont {Wang}},\
  }\bibfield  {title} {\enquote {\bibinfo {title} {{Anderson localization
  transition in a robust $\mathcal{PT}$-symmetric phase of a generalized
  Aubry-Andr\'e model}},}\ }\href {\doibase 10.1103/PhysRevA.103.L011302}
  {\bibfield  {journal} {\bibinfo  {journal} {Phys. Rev. A}\ }\textbf {\bibinfo
  {volume} {103}},\ \bibinfo {pages} {L011302} (\bibinfo {year}
  {2021})}\BibitemShut {NoStop}%
\end{thebibliography}%

%%%%%%%%%%%%%%%%%%%%%%%%%%%
\end{document}